\documentclass[12pt,aps,prd,preprint,preprintnumbers,showpacs,nofootinbib]{revtex4}

\usepackage[dvips]{graphicx}

\input{epsf.sty}

\def\fo{\hbox{{1}\kern-.25em\hbox{l}}}

\def\slashchar#1{\setbox0=\hbox{$#1$}           
   \dimen0=\wd0                                 
   \setbox1=\hbox{/} \dimen1=\wd1               
   \ifdim\dimen0>\dimen1                        
      \rlap{\hbox to \dimen0{\hfil/\hfil}}      
      #1                                        
   \else                                        
      \rlap{\hbox to \dimen1{\hfil$#1$\hfil}}   
      /                                         
   \fi}                                         %

\def\hide#1{[hidden stuff]}

\def\beq{\begin{equation}}
\def\eeq{\end{equation}}
\def\eq{\end{equation}}
\def\to{\rightarrow}

\def\mEt{\mbox{${\hbox{$E$\kern-0.6em\lower-.1ex\hbox{/}}}_T$}\, } 

\def\bsg{\ifmmode B_d\to X_s\gamma\else $B_d\to X_s\gamma$\fi}
\def\bsglue{\ifmmode B_d\to X_s\, g\else $B_d\to X_s\, g$\fi}
\def\bsll{\ifmmode B_d\to X_s\ell^+\ell^-\else $B_d\to X_s\ell^+\ell^-$\fi}
\def\bstt{\ifmmode B\to X_s\tau^+\tau^-\else $B\to X_s\tau^+\tau^-$\fi}
\def\shat{\ifmmode \hat{s}\else $\hat{s}$\fi}

\def\bphik{\ifmmode B_d\to \phi K_s\else $B_d\to \phi K_s$\fi}
\def\bbarphik{\ifmmode \bar{B_d}\to \phi K_s\else $\bar{B_d}\to \phi K_s$\fi}
\def\bsmix{\ifmmode B_s \bar{B}_s\else $B_s \bar{B}_s$\fi}
\def\bdmix{\ifmmode B_d \bar{B}_d\else $B_d \bar{B}_d$\fi}
\def\bclnu{\ifmmode B_d\to X_c e \bar{\nu}\else $B\to X_c e \bar{\nu}$\fi}
\def\bjpsik{\ifmmode B_d\to J/\psi~K_s\else $B_d\to J/\psi~K_s$\fi}
\def\bsee{\ifmmode B_d\to X_s e^+ e^-\else $B_d\to X_s e^+ e^-$\fi}
\def\bsmumu{\ifmmode B_d\to X_s\mu^+\mu^-\else $B_d\to X_s \mu^+\mu^-$\fi}

\def\EmissT{\not \! \!  E_{T}}

\newcommand{\newc}{\newcommand}

\newc{\asusy}{\delta a^{\rm SUSY}_\mu}
\newc{\lcal}{\int {\cal L}dt}
 
\newc{\LSP}{{\chi^0_1}}
\newc{\stauR}{{\tilde \tau_R}}
\newc{\stau}{{\tilde \tau_1}}
\newc{\mstop}{m_{\tilde{t}}}
\newc{\mHpm}{m_{H^\pm}}
\newc{\gsim}{\lower.7ex\hbox{$\;\stackrel{\textstyle>}{\sim}\;$}}
\newc{\lsim}{\lower.7ex\hbox{$\;\stackrel{\textstyle<}{\sim}\;$}}
\newc{\ie}{{\it i.e.}}          
\newc{\etal}{{\it et al.}}
\newc{\eg}{{\it e.g.}}          
\newc{\kev}{\hbox{\rm\,keV}}            
\newc{\mev}{\hbox{\rm\,MeV}}            
\newc{\gev}{\hbox{\rm\,GeV}}            
\newc{\tev}{\hbox{\rm\,TeV}}
\newc{\xpb}{\hbox{\rm\, pb}}
\newc{\xfb}{\hbox{\rm\, fb}}

\newc{\mtop}{m_t}
\newc{\mbot}{m_b}
\newc{\mz}{m_Z}
\newc{\mw}{M_W}
\newc{\alphasmz}{\alpha_s(m_Z^2)}
\newc{\swsq}{\sin^2\theta_W}
\newc{\tw}{\tan\theta_W}
\newc{\cw}{\cos\theta_W}
\newc{\sw}{\sin\theta_W}
\newc{\BR}{\hbox{\rm BR}}
\newc{\zbb}{Z\to b\bar}
\newc{\Gb}{\Gamma (Z\to b\bar b)}
\newc{\Gh}{\Gamma (Z\to \hbox{\rm hadrons})}
\newc{\rbsm}{R_b^\hbox{\rm sm}}
\newc{\rbsusy}{R_b^\hbox{\rm susy}}
\newc{\drb}{\delta R_b}

\newc{\sgn}{\mbox{sgn}}
\newc{\tbeta}{\tan\beta}
\newc{\uL}{{\tilde u_L}}
\newc{\uR}{{\tilde u_R}}
\newc{\cL}{{\tilde c_L}}
\newc{\cR}{{\tilde c_R}}
\newc{\tL}{{\tilde t_L}}
\newc{\tR}{{\tilde t_R}}
\newc{\dL}{{\tilde d_L}}
\newc{\dR}{{\tilde d_R}}
\newc{\sL}{{\tilde s_L}}
\newc{\sR}{{\tilde s_R}}
\newc{\bL}{{\tilde b_L}}
\newc{\bR}{{\tilde b_R}}
\newc{\eL}{{\tilde e_L}}
\newc{\eR}{{\tilde e_R}}
\newc{\mhp}{m_{H^\pm}}
\newc{\mhalf}{m_{1/2}}
\newc{\emt}{{e/\mu /\tau}}

\newc{\lR}{\tilde{l}_R}
\newc{\lL}{\tilde{l}_L}
\newc{\nL}{\tilde{\nu}_L}
\newc{\na}{\chi^0_1}
\newc{\nb}{\chi^0_2}
\newc{\nc}{\chi^0_3}
\newc{\nd}{\chi^0_4}
\newc{\ca}{\chi^{\pm}_1}
\newc{\cb}{\chi^{\pm}_2}
\newc{\camp}{\chi^\mp_1}
\newc{\cbmp}{\chi^\mp_1}
\newc{\capos}{\chi^{+}_1}
\newc{\caneg}{\chi^{-}_1}
\newc{\phit}{\phi_t}
\newc{\phib}{\varphi_b}
\newc{\phiew}{\phi_{ew}}
\newc{\htz}{h^0_t}
\newc{\hbz}{h^0_b}
\newc{\hewz}{h^0_{ew}}
\newc{\hsmz}{h^0_{sm}}
\newc{\huz}{h^0_u}
\newc{\hsusyz}{h^0_{susy}}

\newcommand{\drawsquare}[2]{\hbox{%
\rule{#2pt}{#1pt}\hskip-#2pt
\rule{#1pt}{#2pt}\hskip-#1pt
\rule[#1pt]{#1pt}{#2pt}}\rule[#1pt]{#2pt}{#2pt}\hskip-#2pt
\rule{#2pt}{#1pt}}

\newc{\Dal}{\drawsquare{7}{0.6}}

\def\dofig#1#2{\epsfxsize=#1\centerline{\epsfbox{#2}}}
\def\dofigs#1#2#3{\centerline{\epsfxsize=#1\epsfbox{#2}%
   \hfil\epsfxsize=#1\epsfbox{#3}}}

\def\beq{\begin{equation}}
\def\eeq{\end{equation}}
\def\bea{\begin{eqnarray}}
\def\eea{\end{eqnarray}}

\catcode`@=11
\long\def\@caption#1[#2]#3{\par\addcontentsline{\csname
  ext@#1\endcsname}{#1}{\protect\numberline{\csname
  the#1\endcsname}{\ignorespaces #2}}\begingroup
    \small
    \@parboxrestore
    \@makecaption{\csname fnum@#1\endcsname}{\ignorespaces #3}\par
  \endgroup}
\catcode`@=12

\begin{document}

\begin{flushright}
MSUHEP-040520 \\
hep-ph/0405220
\end{flushright}

\title{
Collider Signature of Bulk Neutrinos in Large Extra Dimensions}

\author{Qing-Hong Cao} \email{cao@pa.msu.edu}
\author{Shrihari Gopalakrishna} \email{shri@pa.msu.edu}
\author{C.-P. Yuan} \email{yuan@pa.msu.edu}

\affiliation{
\vspace*{2mm}
{Department of Physics and Astronomy, \\ 
   Michigan State University, \\
     East Lansing, MI 48824, USA. \\ }}

\date{\today}

\begin{abstract}
We consider the collider signature of right-handed neutrinos propagating in $\delta$ (large) 
extra dimensions, and interacting with Standard Model fields only through a Yukawa coupling 
to the left-handed neutrino and the Higgs boson.
These theories are attractive as they can explain the smallness of the neutrino mass, as
has already been shown. We show that if $\delta$ is bigger than two, it can result in an 
enhancement in the production rate of the Higgs boson, decaying either invisibly or to 
a $b$ anti-$b$ quark pair, associated with an isolated high $p_T$ charged lepton 
and missing transverse energy at future hadron colliders, such as the LHC. 
The enhancement is due to the large number of Kaluza-Klein neutrinos produced in the final state.
The observation of the signal event would provide an opportunity to distinguish between the
normal and inverted neutrino mass hierarchies, and to determine the absolute scale 
of neutrino masses by measuring the asymmetry of the observed event numbers in the electron
and muon channels.  

\end{abstract}


\maketitle

\section{Introduction}
The Standard Model (SM) of high energy physics suffers from the gauge hierarchy problem, 
which is the fine tuning required to maintain a low electroweak scale ($M_{EW} \sim 10^3$~GeV)
in the presence of another seemingly fundamental scale, the Planck scale (the scale of gravity, 
$M_{pl}\sim 10^{19}$~GeV).
Supersymmetry, technicolor and more recently extra (space) dimensions have been proposed to 
address the hierarchy problem. 

Recent neutrino oscillation experiments have suggested a nonzero neutrino mass, with the best
fit values of the mass differences and mixing angles given 
by~\cite{Fukuda:1998mi,Ahmad:2002ka,Wolfenstein:1977ue,Bahcall:2002hv}
\footnote{In this work, we will not address the Liquid Scintillator Neutrino Detector (LSND) 
result~\cite{Aguilar:2001ty}.}
\bea
\Delta m^2_{\rm solar} &=& 7\times 10^{-5}~{\rm eV^2} \ , \ \ \ \tan^2\theta_{solar} = 0.4 \ , \nonumber \\
\Delta m^2_{\rm atm} &=& 2.5\times 10^{-3}~{\rm eV^2} \ , \ \ \ \tan^2\theta_{atm} = 1 \ .
\label{DMSQ.EQ}
\eea
The oscillation between the three active flavors,  $\nu_e$,$\nu_\mu$,$\nu_\tau$, in the SM, 
accommodates this satisfactorily, with 
$\Delta m^2_{\rm solar} = (m_2^{2}-m_1^{2})$ and $\Delta m^2_{\rm atm} = |m_2^{2}-m_3^{2}|$, where the 
$m_i$ are the physical neutrino masses.
If the $m_i$ are also assumed to be of the same order of magnitude as the mass differences, 
it is quite challenging to explain why it is that the neutrinos are so light compared to the 
other leptons. 

It has been shown~\cite{Arkani-Hamed:1998rs,Antoniadis:1998ig} that if 
there are other Large Extra Dimensions (LED) in addition to our usual four space-time dimensions, 
we could potentially solve the gauge hierarchy problem. It was then pointed 
out~\cite{Dienes:1998sb,Arkani-Hamed:1998vp,Dvali:1999cn} that the smallness of the neutrino 
mass is naturally explained\footnote{We note here that the 
conventional see-saw mechanism to explain the smallness of 
the neutrino mass is equally appealing, but we will not consider it in this work.} 
if right-handed neutrinos that propagate in some $\delta$ number of these extra 
dimensions are introduced. We will refer to such neutrinos, which are SM gauge singlets, as 
``bulk neutrinos'', as is the usual practice. Various aspects of theories with bulk neutrinos have been
analyzed in Ref.~\cite{Mohapatra:1999zd}. 

We will take the view, as in Ref.~\cite{Davoudiasl:2002fq}, that the standard three-active-flavor 
oscillation explains the data in Eq.~(\ref{DMSQ.EQ}), and that 
the mixing to sterile bulk neutrinos are small enough to evade experimental constraints. 
In our earlier work~\cite{Cao:2003yx}, we considered the constraints on theories with bulk neutrinos
coming from neutrino oscillation experiments and also from requiring that perturbative
unitarity be maintained in the theory in Higgs-Higgs scattering. We showed that 
strong constraints result when $\delta > 2$, though in that case, a precise calculation of 
the bound was not possible owing to the sensitivity on the cutoff scale, implying a 
dependence on the completion of the extra dimensional (effective) theory that we work with. 
There, we noted that as pointed out in Ref.~\cite{Davoudiasl:2002fq}, an alternative 
approach~\cite{Dienes:1998sb,Dvali:1999cn} wherein Eq.~(\ref{DMSQ.EQ}) is explained by the 
oscillation of the active species predominantly into sterile bulk neutrinos, 
appears to be disfavored by the Sudbury Neutrino Observatory~(SNO)~\cite{Ahmad:2002ka} 
neutral current data. 

If bulk right-handed neutrinos are responsible for the smallness of the 
neutrino mass, we ask in this paper, what the consequences at a collider might be. 
We will show that in certain cases, the Higgs production and decay could be 
measurably altered from the SM expectation. We will present the discovery potential of hadron 
colliders, mainly focusing on the CERN Large Hadron Collider (LHC), through the mode: 
$q q^\prime \rightarrow W \rightarrow {\rm Higgs + lepton + \EmissT}$. 
The invisible decay mode of the Higgs boson in theories with a bulk right-handed neutrino 
at hadron colliders have also been explored in Refs.~\cite{Martin:1999qf,Godbole:2003it}.

The rest of the paper is organized as follows. 
We will introduce the extra dimensional theory with bulk neutrinos in Sec.~\ref{BNLED.SEC}, 
write down the equivalent four dimensional Kaluza-Klein theory, with particular focus on
the interaction of the right-handed bulk neutrino with the Higgs field and the left-handed
neutrino. In Sec.~\ref{COLSIG.SEC}, we will analyze the collider signature of such a theory at the 
Fermilab Tevatron and the CERN LHC by performing a Monte-Carlo simulation, 
and improve the significance by appropriate cuts. Our conclusions are given in Sec.~\ref{CONCL.SEC}. 

\section{Right-handed Neutrinos in Extra Dimensions}
\label{BNLED.SEC}
In this section, we summarize the framework with bulk right-handed neutrinos, details of
which are given in Ref.~\cite{Cao:2003yx}.

To address the gauge hierarchy problem, Arkani-Hamed, Dimopoulos and Dvali 
(ADD)~\cite{Arkani-Hamed:1998rs} postulate that the 
Standard Model (SM) fields are confined to a 4 dimensional (4-D) sub-space (brane) in an 
extra dimensional world of $4+n$ total dimensions. 
ADD take the view that the only fundamental scale in nature is $M_*$, which is of the order of 
$M_{EW}$, and the apparent 4-D gravity scale ($M_{pl}$) is then given by
\bea
M_{pl}^2 = M_*^{2+n} V_n,
\label{ADDMPL.EQ}
\eea
where, $V_n$ is the volume of the (compact) extra dimensional space. In the simple case of 
each of the compact extra dimensions being of equal radius $R^\prime$, we have $V_n \sim {R^\prime}^n$. 
Thus ADD argue that $M_{pl}$ appears to be a large scale from a 4 dimensional perspective 
simply because the volume $V_n$ is large. In other words, the explanation of why $M_{pl}$ is 
large is recast to stabilizing $R^\prime$ at a large value, so that $V_n$ is large.

It should be pointed out that for a given $M_*$, if the $n$ compact dimensions have equal 
radii $R^\prime$, Eq.~(\ref{ADDMPL.EQ}) implies a
particular value of $R^\prime$. However, if it happens that there are two sets of compact extra 
dimensions of unequal size, $\delta$ of them ($\delta \leq n$) with radius $R$, and the other
$(n-\delta)$ with radius $R^\prime$,
then we have in this case, $V_n \sim  {R^\prime}^{(n-\delta)}\,R^{\delta}$. We can in this case
think of $R$ as an independent variable with $R^\prime$ being determined by 
Eq.~(\ref{ADDMPL.EQ}).  

We consider the ADD framework, to which is added three (one for each generation) bulk fermions, 
$\Psi^\alpha (x^\mu,\underline{y})$, that propagate in 4+$\delta$ dimensions 
($\delta$ of them compact with radius $R$), where the indices 
$\alpha$,~$\beta=(1,~2,~3)$ denote the three generations, and $\underline{y}$ 
stands for $\{y^1,...,y^\delta\}$. 

\subsection{The Lagrangian} 
We can split the Lagrangian into a bulk piece and a brane piece,
\bea
{\cal S} &=& \int d^4x~ d^\delta y~\left[{\cal L}_{\rm Bulk}~+~\delta(\underline y)\,
{\cal L}_{Brane}\right].
\eea
${\cal L}_{\rm Bulk}$ contains the Einstein-Hilbert bulk gravity term (which we will not 
show explicitly, but can be found, for example, in Ref.~\cite{Giudice:1998ck}),
the kinetic energy term for the bulk neutrino field $\Psi(x^\mu,\underline{y})$ and in 
general, a bulk Majorana mass term for $\Psi$, which for simplicity we will omit 
(see Ref.~\cite{Dienes:1998sb} for implications of a nonzero bulk Majorana mass).   
${\cal L}_{\rm Brane}$ contains the SM Lagrangian plus an interaction term between SM fields 
and $\psi_R$,
\bea
{\cal L}_{\rm Bulk} &\supset& \bar\Psi^\alpha\, i\Gamma^M D_M \Psi^\alpha, \nonumber \\
{\cal L}_{\rm Brane} &\supset& {\cal L}_{\rm SM} - \left(\frac{\Lambda^\nu_{\alpha\beta}}
{\sqrt{M_*^\delta}}~h\,\psi_R^\beta\,\nu_L^\alpha + h.c. \right),  \label{BULKL.EQ} \\  
{\cal L}_{\rm SM} &\supset& \bar\nu_L^\alpha\, i\gamma^\mu D_\mu \nu_L^\alpha + 
\left(\frac{g}{\sqrt{2}}\,\bar\nu_L^\alpha\, \gamma^\mu e_L^\alpha W_\mu^+ + h.c\right) 
+ ...\, , \nonumber
\eea
where, $\Lambda^\nu_{\alpha\beta}$ is an $O(1)$ Yukawa coupling constant. It should be 
kept in mind that $\psi_R$ is a function of ($x^\mu,\underline y$) whereas the SM fields 
are functions of $x^\mu$ only. The index $M$ runs over \{$x^\mu,\underline y$\}.

We can perform a Kaluza-Klein (KK) expansion of the 4+$\delta$ dimensional theory and obtain 
an equivalent 4-D theory by writing, 
\bea
\psi^\alpha_{R}(x^\mu,\underline{y}) = \sum_{\underline n}~
{\psi_R^\alpha}^{(\underline n)}(x^\mu)~f_{\underline n}(\underline{y}),
\eea
where, $\underline n = (n_1,...,n_\delta)$ is a vector in ``number space'', 
$\psi^{(\underline n)}$ are the KK modes and $f_{\underline n}(\underline{y})$ is a 
complete set over $\underline{y}$. A similar expansion is made for $\psi^\alpha_{L}$. 
To reduce clutter, we will simply write $n$ and $y$ for $\underline n$ and $\underline y$, 
respectively. We will use the notation $n=(0,1,...)$ and $\hat n=(1,...)$ ($\hat n$ excludes 0). 
$f_n$ is an orthonormal set,
\bea
\int_{0}^{2\pi R} d^\delta y ~~ f_n^*(y) f_{m}(y) = \delta^{n m},
\eea 
and a convenient choice is,
\bea
f_{n}(y) = \frac{e^{i \frac{n.y}{R}}}{\sqrt{\frac{S_{\delta-1}}{\delta} R^\delta}} = 
\frac{e^{i \frac{n.y}{R}}}{\sqrt{V_\delta}},
\eea
with $S_{\delta-1}$ the surface ``area'' of a unit sphere in $\delta$ dimensions, and 
$V_\delta \equiv \frac{S_{\delta-1}}{\delta} R^\delta$ is the volume of the extra dimensional space.

We define the fields\footnote{The other linear combinations 
$\left(\psi_R^{\alpha (\hat n)} - \psi_R^{\alpha (-\hat n)} \right)$ and 
$\left(\psi_L^{\alpha (\hat n)} - \psi_L^{\alpha (-\hat n)} \right)$ are decoupled from the SM fields, 
and we will not consider them further. Also, with orbifold compactification, we can project out
$\psi_L^{\alpha (0)}$ so that it is excluded from the particle 
spectrum.}~\cite{Davoudiasl:2002fq},
\bea
\nu^\alpha_R &\equiv& \psi_R^{\alpha (0)}, \nonumber \\
\nu_R^{\alpha (\hat n)} &\equiv& \frac{1}{\sqrt{2}} \left(\psi_R^{\alpha (\hat n)} + 
\psi_R^{\alpha (-\hat n)} \right),  \nonumber \\
\nu_L^{\alpha (\hat n)} &\equiv& \frac{1}{\sqrt{2}} \left(\psi_L^{\alpha (\hat n)} + 
\psi_L^{\alpha (-\hat n)} \right) \ . \nonumber
\eea
We substitute the KK expansion for the bulk fields $\psi_R^\alpha$ and 
$\psi_L^\alpha$ into Eq.~(\ref{BULKL.EQ}) to get the equivalent 4-D theory,
\beq
{\cal L}^{(4)} = {\cal L}_{SM} - \sum_{\alpha=1}^{3}\sum_{\hat n} \left[ \frac{|\hat n|}{R}
\left(\nu_R^{\alpha (\hat n)} \nu_L^{\alpha {(\hat n)}} + h.c.\right) \right] -
\sum_{\alpha,\beta=1}^{3} \left[\frac{m_\nu^{\alpha\beta}}{v} \left(h \nu_R^\alpha \nu_L^\beta + 
\sum_{\hat n} \sqrt{2} h \nu_R^{\alpha (\hat n)} \nu_L^\beta \right) + h.c.\right] \ ,
\label{EQ4DL1.EQ}
\eeq 
where, $|\hat n|~\equiv~\sqrt{n_1^2+...+n_\delta^2}$. We note here that in ${\cal L}^{(4)}$, there is 
a tower of KK states $(\nu_L^{(\hat n)},\nu_R^{(\hat n)})$ with Dirac masses approximately equal to 
$|\hat n|/R$.
With $SU(2)$ broken by the Higgs mechanism, 
by the Higgs field acquiring a vacuum expectation value (VEV), $\left<h\right>=v$, we have 
the neutrino mass matrix given by,
\bea
m_\nu^{\alpha \beta} \equiv 
\frac{\Lambda^\nu_{\alpha \beta}~v}{\sqrt{\frac{S_{\delta-1}}{\delta}(M_*R)^\delta}} = 
\frac{\Lambda^\nu_{\alpha \beta}~v}{\sqrt{V_\delta M_*^\delta}},
\eea
and $m_\nu^{\alpha\beta}$ can be much smaller than $v \sim M_{EW}$, if $V_\delta$ is large
for $\Lambda^\nu_{\alpha \beta}$ being $O(1)$.
Henceforth, we will assume that unless noted otherwise,
repeated generation indices $\alpha,\beta,i,j$, and KK indices $n,\hat n,m$, are summed over.

\subsection{KK states}
\label{KKD.SEC}
For $\delta > 1$, the state with mass 
$\frac{|\hat n|}{R}$ can be degenerate, and we denote the degeneracy at the $\hat n^{\rm th}$ 
level by $d_{\hat n}$. (Strictly speaking, we should denote this as $d_{|\hat n|}$, but we will 
just write this as $d_{\hat n}$.) For example, for $\delta=3$, the state with mass 
$1/R$ has $d_1 = 3$, corresponding to 
$(\hat n_1,\hat n_2,\hat n_3)\rightarrow (1,0,0),\ (0,1,0)\ {\rm and}\ (0,0,1)$, all
of which have the same mass. For large $|\hat n|$, the leading power dependence of $d_{\hat n}$
in $\delta$ extra dimensions is given by $d_{\hat n} = c_{\hat n} |\hat n|^{\delta - 1}$, 
where the $c_{\hat n}$ are $O(1)$ numbers. We define $d_0 \equiv 1$.

For large $|\hat n|$, we can think of the $\hat n_i$ as a continuum and the leading behavior 
is given by the surface of the $(\delta-1)$-sphere of radius $|\hat n|$ in number space, 
\bea
d_{\hat n} \sim S_{\delta-1} |\hat n|^{\delta-1} \ \ \ \ \ \ {\rm (in\ \delta\ dimensions).} 
\label{DNC.EQ} \\
d^{\delta = 1}_{\hat n} = 1, \qquad d^{\delta = 2}_{\hat n} \sim 2\pi\hat n, \qquad
d^{\delta = 3}_{\hat n} \sim 4\pi\hat n^2 \ . \nonumber
\eea
For example, for $\delta=3$, $d_{\hat n} \sim 4\pi |\hat n|^2$, which is the surface of 
the 2-sphere with radius $|\hat n|$. We will often use the continuum approximation for 
estimating various quantities.

In a collider, the heaviest KK state that could be produced in the final state is limited by the 
center-of-mass energy $\sqrt{s}$. We define $N_s$ to be the radius of the biggest sphere in 
$\{n_i\}$ space such that $N_s/R = \sqrt{s}$. The sum over the KK states of certain quantities 
can be divergent and can depend on $N_s$. We will elaborate more on this later, and we will 
see that the production rate of the KK states in association with a Higgs boson can depend on $N_s$, 
especially strongly for $\delta>3$.

\subsection{Higgs boson interaction}
\label{BNLED_HINT.SEC}
We can make the Yukawa coupling in Eq.~(\ref{EQ4DL1.EQ}) diagonal in generation space 
($\alpha, \beta$) with the rotations~\cite{Davoudiasl:2002fq},
\bea
\nu_L^\alpha = l^{\alpha i} \nu_L^{\prime i} &,& \ \ \
\nu_R^\alpha = (r^{\alpha i})^* \nu_R^{\prime i} \ , \nonumber \\
\nu_R^{\alpha (\hat n)} = (r^{\alpha i})^* \nu_R^{\prime i (\hat n)} &,& \ \ \
\nu_L^{\alpha (\hat n)} = r^{\alpha i} \nu_L^{\prime i (\hat n)} \ , \nonumber \\
e_L^\alpha = l_e^{\alpha i} e_L^{\prime i} &,& \ \ \
e_R^\alpha = (r_e^{\alpha i})^* e_R^{\prime i} \ , \nonumber
\label{MNSROT.EQ}
\eea
 where the unitary matrices $l$ and $r$ are chosen to diagonalize $m_\nu^{\alpha\beta}$, so that, 
$(r^{\alpha i})^*~m_\nu^{\alpha\beta}~(l^{\beta j})=m_i \delta^{i j}$. Similarly, $l_e$ 
and $r_e$ are chosen to diagonalize the electron-type mass matrix. 
In the usual way, the charged current interactions now become proportional to the MNS 
matrix~\cite{Maki:mu},
\beq
V_{MNS} \equiv l_e^\dag\, l \ .
\label{MNSDEF.EQ}
\eeq
We can work in the basis in which the charged lepton mass matrix is diagonal. In this case we have from
Eq.~(\ref{MNSDEF.EQ}), $V_{MNS} = l$. Based on the $V_{MNS}$ obtained 
from a global fit to the oscillation data, we take the $l$ to be such that the solar neutrino 
mixing angle between $\nu_e~\&~\nu_\mu$ is given by $\tan^2\theta_{e\mu}=0.4$, and the 
atmospheric oscillation mixing between $\nu_\mu~\&~\nu_\tau $ is maximal.\footnote{For simplicity,
we assume the small mixing angle $\theta_{13}\approx 0$, which leads to $l^{e3}\approx 0$.}
This implies
\beq
l = \pmatrix{
0.847 & 0.531 & 0 \cr
-0.376 & 0.599 & 0.707\cr
0.375 & -0.599 & 0.707
}.
\label{lBFIT.EQ}
\eeq

We take the value of $\Lambda^\nu_{\alpha \beta}$  such that the lowest mass eigenvalue
$m_i$ obtained by diagonalizing the neutrino mass matrix~\cite{Cao:2003yx} is consistent 
with Eq.~(\ref{DMSQ.EQ}). 
Defining $\xi_i \equiv m_i R$, we showed in Ref.~\cite{Cao:2003yx} that in order to satisfy 
the limits on the probability of an active neutrino oscillating into the sterile KK states, 
we are led to restrict ourselves to the case when 
$\sum_{\hat n} \frac{\xi_i^2}{\hat n^2} d_{\hat n} \ll 1$ (for all $\delta$). 

We will show in Sec.~\ref{COLSIG_PROD.SEC} that the production rate of the KK neutrinos 
in association with the Higgs boson is proportional to the matrix
\beq
\bar{m}_\nu \equiv m_0 l^\dagger \ ,  
\eeq
with $m_0$ the diagonal $3\times 3$~physical neutrino mass matrix $m_i \delta^{ij}$. 
We will find it convenient to work in a basis in which we write the Higgs interaction 
term in Eq.~(\ref{EQ4DL1.EQ}) as
\beq
{\cal L}^{(4)} \supset - \left[\frac{\bar{m}_\nu^{i\beta}}{v} 
\left(h \nu_R^{\prime i} \nu_L^\beta + 
\sum_{\hat n} \sqrt{2} h \nu_R^{\prime i(\hat n)} \nu_L^\beta \right) + h.c.\right] \ ,
\label{EQ4DLRM.EQ}
\eeq
where we have rotated the $\nu_R^\alpha$ with the matrix $r$ while retaining the $\nu_L$ 
in the flavor basis, in order to explicitly keep only the physical matrix $l$, while 
absorbing the unphysical matrix $r$ into the definition of $\nu_R^\prime$. In particular,
we will also show in Sec.~\ref{COLSIG_PROD.SEC} that the total production rate 
(sum over $e,~\mu~{\rm and}~\tau$ channels) of our signal process is proportional to
$\sum_{i,\ell} |\bar{m}_{\nu}^{i\ell}|^2$.  

\begin{figure}
\dofig{4in}{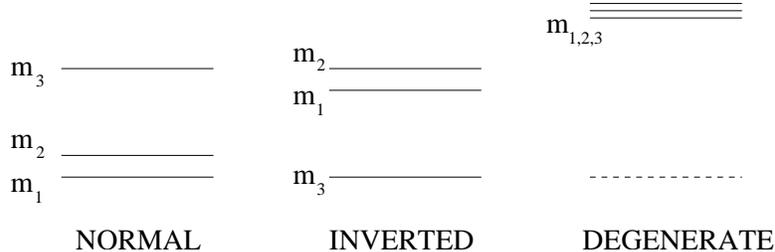}
\caption{Neutrino mass schemes. 
\label{MNUSCH.EQ}}
\end{figure}

Presently the combination of various neutrino oscillation experiments constrain 
$|\Delta m^2_{\rm atm}|$ leaving its sign undetermined, while the solar 
MSW effect~\cite{Wolfenstein:1977ue} fixes the sign and magnitude of $\Delta m^2_{\rm solar}$. 
Furthermore, oscillation experiments are only sensitive to the mass differences due to which
the absolute scale of neutrino masses is not known. 
This leads to more than one possibility for the values of the neutrino mass eigen-values, and 
in Ref.~\cite{Cao:2003yx}, we noted the three neutrino mass schemes shown in Fig.~\ref{MNUSCH.EQ}.
In each scheme, we give below the matrix $\bar{m}_\nu$ which specifies the 
Higgs interaction that we use to calculate the collider signature in the next section. 
\begin{itemize}
\item[(i)] Normal Hierarchy: $m_1\approx 0$, $m_2\approx 0.008$~eV and $m_3\approx 0.05$~eV, and  
\bea
\bar{m}_\nu = \pmatrix{
0 & 0 & 0 \cr
0.0045 & 0.005 & -0.005 \cr
0 & 0.036 & 0.036
} \ .
\label{mnlnor.EQ}
\eea
\item[(ii)] Inverted Hierarchy: $m_1,m_2\approx 0.05$~eV and $m_3\approx 0$, and
\bea
\bar{m}_\nu = \pmatrix{
0.043 & -0.019 & 0.019 \cr
0.027 & 0.03 & -0.03 \cr
0 & 0 & 0
} \ .
\label{mnlinv.EQ}
\eea
\item[(iii)] Degenerate: $m_1,m_2,m_3$ all at some mass scale less than the limits discussed in 
Ref.~\cite{Cao:2003yx}. Here, to illustrate
the character of the signature, we take the three masses to be around 1~eV, an arbitrary choice, 
and 
 \bea
\bar{m}_\nu = \pmatrix{
0.847 & -0.376 & 0.375 \cr
0.531 & 0.599 & -0.599 \cr
0 & 0.708 & 0.708 
} \ .
\label{mnldeg.EQ}
\eea
\end{itemize}

As shown in Eq.~(\ref{EQ4DLRM.EQ}), the Higgs boson couples to the tower of KK neutrino states
with the same Yukawa coupling that is proportional to the small neutrino mass. When $1/R$ is 
small and $\delta$ is large, there can be a large number of KK states that may overcome the
small Yukawa coupling and generate an observable collider signature in the production and
the decay of the Higgs boson. We shall discuss this possibility in the next section. 

\section{Collider Signature}
\label{COLSIG.SEC}
\subsection{Production of Higgs boson}
\label{COLSIG_PROD.SEC}
\begin{figure}
\dofig{2.5in}{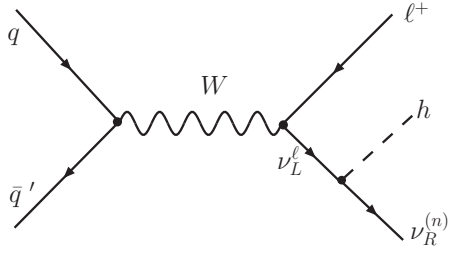}
\caption{Signal process \label{fig:signal}}
\end{figure}

The right-handed neutrino does not carry any electroweak quantum numbers and therefore
can be produced in a collider only through the Yukawa interaction.
Although the production cross section is suppressed by the small Yukawa coupling (since it is
proportional to the small neutrino mass), the cross section could get a large enhancement when 
summed over the large number of allowed KK excitations of the bulk neutrino. 
In such a case, the Higgs phenomenology can be altered in an interesting way and we study this 
in the process:
\bea
q\bar{q}'\rightarrow W^{*}\rightarrow \ell^{+}\, h\, \nu_{R}^{\prime i (n)}, \label{eq:prod_process}
\eea
as shown in Fig.~\ref{fig:signal}, where $\ell^+$ is a charged lepton, $\nu^{\prime i (n)}_R$
the $n$-th KK mass eigen-state and the process is mediated by $\nu_L^{\alpha}$ with $\alpha = \ell$. 
The production of $\nu^{\prime i (n)}_R$ in the final state would lead to missing energy in the event.

The total cross section for this process at a hadron collider is
\bea
\sigma(P_1 P_2\rightarrow \ell^+ h \nu^{(n)}_{R})&=& 
   \sum_{n} \sigma^{(n)}(P_1 P_2\rightarrow \ell^+ h \nu^{(n)}_{R}) \, d_n \label{eq:hadron_cs} \\ 
 &=&\sum_{n} \sum_{q,\bar{q}^\prime}\int {\rm d}x_1 {\rm d}x_2
 \Biggl[  
     f_{q/P_1}(x_1,\mu) f_{\bar{q}^\prime/P_2}(x_2,\mu) 
     \hat{\sigma}^{(n)} (q \bar{q}^\prime \rightarrow \ell^+ h \nu^{(n)}_{R}) \Biggl. \nonumber \\    
   &&~~~~~~~~~~~~~~~~~~~~~~~~~~~~~~~~~~~~~~~~~~~~  \Biggl. +(x_1\leftrightarrow x_2) 
 \Biggl] \, d_n, \nonumber
\eea
where $\sum_n$ is a sum over the allowed KK states (0 up to $N_s \equiv \sqrt{s} R$) 
with degeneracy $d_n$ at the $n^{\rm th}$ level, 
$P_1,P_2$ represent the hadronic initial state,
$f_{q/P}(x,\mu)$ is the parton distribution function (PDF). CTEQ6\,M~\cite{Pumplin:2002vw} 
is adopted in our calculation. We take the factorization scale $\mu$ to be the invariant mass
of the constituent process in our numerical calculation. $\hat{\sigma}^{(n)}$ is the parton 
level cross section to produce the $n^{\rm th}$ KK mass eigen-state $\nu^{\prime i (n)}_{R}$, 
and is given by
\bea
  \hat{\sigma}^{(n)}=\frac{1}{2\hat{s}}\int {\rm d}\Pi_{3}\sum_{i} \sum_{{\rm spin}\atop{\rm color}}
  \overline{\left|{\mathcal M}^{(n)} (q \bar{q}^\prime \rightarrow \ell^+ h \nu^{\prime i (n)}_{R})\right|^2}\ , 
  \label{eq:parton_cs}
\eea
where the bar above the $\left|{\mathcal M}^{(n)}\right|^2$ denotes averaging over the initial
state spin and color, ${\rm d}\Pi_3$ represents 3-body final state phase space, and
the squared matrix element resulting from Eq.~(\ref{EQ4DLRM.EQ}) is given by
\bea
\overline{\left|{\mathcal M}^{(n)}\right|^{2}} & = & 
\left(\,\frac{1}{2}~\frac{1}{2}\,\right)\,\frac{1}{3}\,
\left[\sqrt{2}-(\sqrt{2}-1)\delta_{n,0}\right]^{2} \left(\frac{g_{W}}{\sqrt{2}}\right)^{2}
 \frac{\left|\bar{m}_{\nu}^{i\ell}\right|^2}{v^2}
\frac{1}{(p^{2}-m_{W}^{2})^{2}(q^{2}-m_{\nu}^{2})^{2}} \nonumber \\ 
 &\times& \Biggl{\{}32(p_{\ell^{+}}\cdot p_{q})(p_{h}\cdot p_{\nu_{R}^{(n)}})(p_{h}\cdot p_{\bar{q}^\prime})+
 32m_{\nu_{R}^{(n)}}^{2}(p_{\ell^{+}}\cdot p_{q})(p_{h}\cdot p_{\bar{q}^\prime}) \Biggl. \nonumber \\
 &&~~\Biggl.  + 16m_{\nu_{R}^{(n)}}^{2}(p_{\ell^{+}}\cdot p_{q})(p_{\nu_{R}^{(n)}}\cdot p_{\bar{q}^\prime}) - 16m_{h}^{2}(p_{\ell^{+}}\cdot p_{q})(p_{\nu_{R}^{(n)}}\cdot p_{\bar{q}^\prime}) \Biggl{\}} , 
  \label{eq:parton_msq} 
\eea
where $p_{q}$ and $p_{\bar{q}}$ are the 4-momenta of the incoming partons $q$ and $\bar q$, 
$p_{\ell^{+}}$, $p_{h}$ and $p_{\nu_{R}^{(n)}}$ are the 4-momenta of the outgoing particles, 
$p=p_{q}+p_{\bar{q}^\prime}$ is the momentum of the virtual $W$ boson and 
$q=p_{h}+p_{\nu_{R}^{(n)}}$ is the momentum of the intermediate $\nu^\ell_L$. 
The factor $\left(\frac{1}{2}~\frac{1}{2}\right)$ is due to the spin averaging, 
$\frac{1}{3}$ is the color factor, and $v(=246~{\rm GeV})$ is the usual Higgs field VEV. 

\begin{figure}
\begin{center}
\scalebox{0.65}{\includegraphics{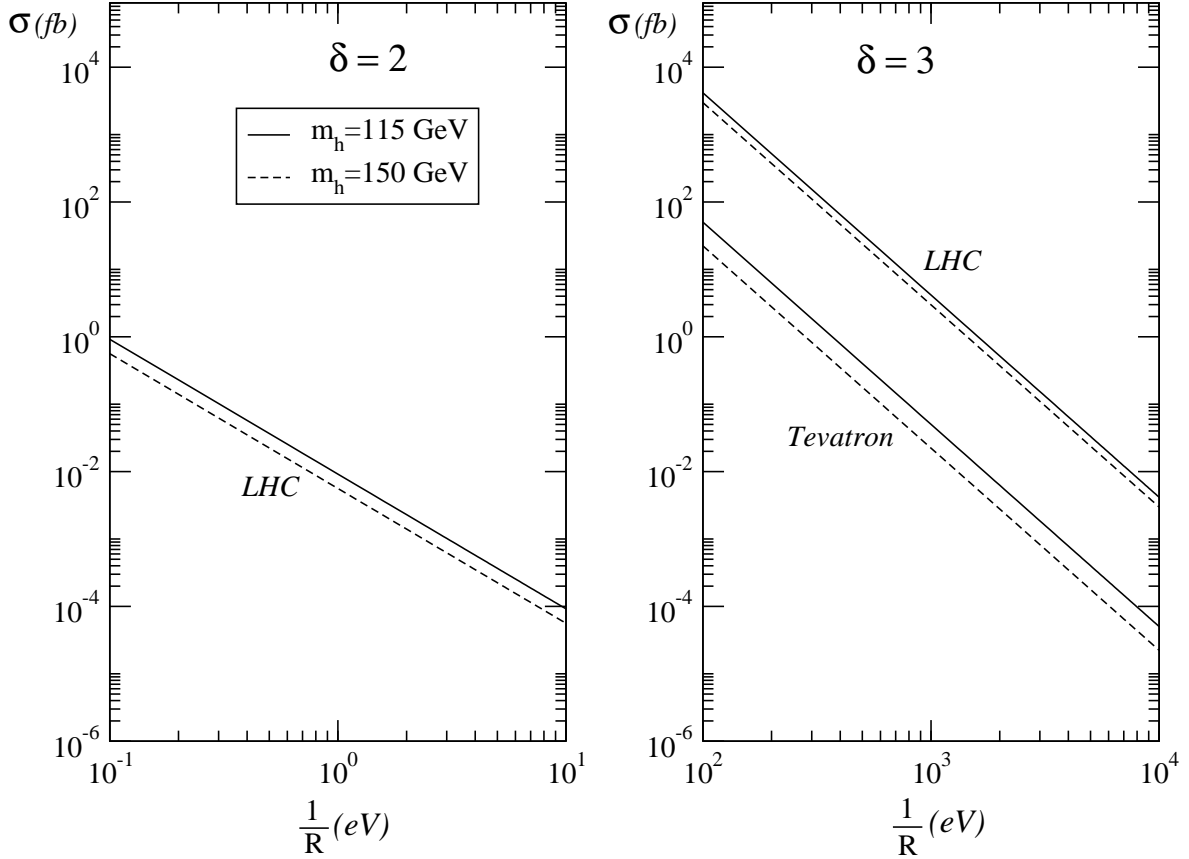}}
\caption{Total cross section (with no cuts applied) as a function of $1/R$ for $m_h=115 {\rm GeV}$ and
         $m_h=150 {\rm GeV}$ at the Tevatron and the LHC for $\sum_{i,\ell} |\bar{m}_{\nu}^{i\ell}|^2 = (0.05~{\rm eV})^2$.
\label{fig:prod_cs}}
\end{center}
\end{figure}

In the continuum approximation (c.f. Eq.~(\ref{DNC.EQ})) after summing over the allowed KK states, 
the total cross section (including $e^+,\mu^+~{\rm and}~\tau^+$ channels) scales as
\bea
\sigma \propto \sum_{i,\ell} \frac{|\bar{m}_{\nu}^{i\ell}|^2}{(1/R)^{\delta}}. \label{eq:scaling}
\label{SIGSCA.EQ}
\eea
Using Eqs.~(\ref{eq:hadron_cs})-(\ref{eq:parton_msq}) we compute the production cross section
of the Higgs boson for different choices of $m_h$, $1/R$ and $\delta$ at the Fermilab Tevatron
($p\bar{p}$ collision at $\sqrt{s}=1.96$ TeV) and the CERN LHC
($pp$ collision at $\sqrt{s}=14$ TeV).
The sum of the Higgs boson production cross section 
(including $\ell^+ = e^+,\mu^+,\tau^+$) is shown in Fig.~\ref{fig:prod_cs} and Table~\ref{tbl:prod-cs},
for a representative choice of $\sum_{i,\ell} |\bar{m}_{\nu}^{i\ell}|^2 = (0.05~{\rm eV})^2$.
(The actual numbers for a particular mass matrix can be obtained easily using the scaling 
relation shown in Eq.~(\ref{SIGSCA.EQ}).)
Although we do not strictly impose the constraints on $1/R$ that we derived in our early
work~\cite{Cao:2003yx}, we choose the range of $1/R$ keeping this in mind.
For this range of $1/R$, we see from Table~\ref{tbl:prod-cs} that the signal rate would be too
small to be seen at the Tevatron, and we therefore only consider the LHC in the rest of this work.
Due to the summation over the KK states in Eq.~(\ref{eq:hadron_cs}), the production cross section
increases with decreasing $1/R$, owing to the increase in the number of KK states
(c.f. Sec.~\ref{KKD.SEC}) that can be produced as final state particles.
Also, as the absolute neutrino masses $m_i$ increase, the $\bar{m}_\nu^{i\ell}$ increase
(for example see Eq.~(\ref{mnldeg.EQ}) for the degenerate mass scheme) and the total cross section
increases with the dependence shown in Eq.~(\ref{SIGSCA.EQ}). However, as the absolute neutrino masses
increase, the constraints on $1/R$ from neutrino oscillation data and from unitarity becomes 
stronger, and imposing the stronger constraint keeps the total cross section from increasing. 

In what we call the degenerate mass scheme, we arbitrarily pick the neutrino mass scale to be
around $1$~eV. Using the scaling relation in Eq.~(\ref{eq:scaling}), one can easily get the 
total cross section for the degenerate mass scheme from Fig.~\ref{fig:prod_cs} and
Table~\ref{tbl:prod-cs}. In the following, when we present results for the degenerate case,
we keep in mind the stronger constraints on $1/R$ (see Ref.~\cite{Cao:2003yx}), and choose 
$1/R$ such that we obtain the same order of magnitude cross section as for the normal and 
inverted mass schemes. 

\begin{table}
\begin{center}
\caption{Cross section (in fb, with no cuts applied) of the process 
         $q\bar{q}^\prime\rightarrow \ell^{+}h\nu_R $, with $\ell^+=e^+,\mu^+,\tau^+$, 
         at the Tevatron and the LHC for different choices of $m_h$, $1/R$ and $\delta$,
         for $\sum_{i,\ell} |\bar{m}_{\nu}^{i\ell}|^2 = (0.05~{\rm eV})^2$    .
         \label{tbl:prod-cs} }
\begin{tabular}{c|c|c|c|c|c}
\hline 
\hline
&
$1/R$&
\multicolumn{2}{c|}{$m_{h}=115$ GeV}&
\multicolumn{2}{c}{$m_{h}=150$ GeV}\tabularnewline
\cline{3-4} \cline{5-6} 
&
(eV)&
Tevatron&
LHC&
Tevatron&
LHC\tabularnewline
\hline
&
$10^{2}$&
$49.2$&
$4.03\times10^{3}$&
$21.9$&
$2.92\times10^{3}$\tabularnewline
$\delta=3$&
$10^{3}$&
$48.2\times10^{-3}$&
$4.03$&
$21.9\times10^{-3}$&
$2.92$\tabularnewline
&
$10^{4}$&
$49.2\times10^{-6}$&
$4.03\times10^{-3}$&
$21.9\times10^{-6}$&
$2.92\times10^{-3}$\tabularnewline
\hline
$\delta=2$&
$1$&
$0.23\times10^{-3}$&
$0.009$&
$0.093\times10^{-3}$&
$0.005$\tabularnewline
\hline
\hline
\end{tabular}
\end{center}
\end{table}

We will study the decay modes of the Higgs boson in order to identify the collider signature of 
the signal events. In addition to the decay modes present in the SM, the Higgs boson can also decay
into ($\nu_L~+~\nu_R^{(n)})$, an invisible decay mode. We will study this next.  

\subsection{Decay of Higgs boson}
The invisible decay width (summed over all neutrino flavors) of the Higgs boson
$(h\rightarrow \nu_{L}\nu_{R}^{(n)})$ is given by
\bea
\Gamma_{\rm invis} & = & \sum_{n}\Gamma_{n} d_n = \sum_{i,\ell} \sum_{n}^{N_{m}}\frac{1}{4\pi}
\left| \frac{\bar{m}_\nu^{i\ell}}{v}\right|^{2}\,m_{h}\left(1-\frac{m_{\nu_{R}^{(n)}}^{2}}{m_{h}^{2}}\right)^{2} d_n ,
\eea
where the sum over the KK states is up to $N_m\equiv m_h R$ since the most massive KK state
that the Higgs boson could decay into is limited by $m_h$. Using the continuum approximation 
to perform the sum over the KK states (c.f. Eq.~(\ref{DNC.EQ})) we get for the invisible 
decay width of the Higgs boson, 
\bea
\Gamma_{\rm invis} =
\left\{ 
 \begin{array}{c}  
     \displaystyle \sum_{i,\ell} \frac{1}{12}m_{h}\,
      \left|\frac{\bar{m}_\nu^{i\ell}}{v}\right|^{2}\,\left(m_{h}R\right)^{2},\qquad\;(\delta=2)\\ [3mm]
     \displaystyle	
     \sum_{i,\ell} \frac{8}{105}m_{h}\,\left|\frac{\bar{m}_\nu^{i\ell}}{v}\right|^{2}
 \,\left(m_{h}R\right)^{3}.\qquad(\delta=3)
 \end{array}
\right. 
\eea

\begin{table}
\begin{center}
\caption{Decay branching ratio of the Higgs boson for different choices of $m_h$, $1/R$ and $\delta$,
        for $\sum_{i,\ell} |\bar{m}_{\nu}^{i\ell}|^2 = (0.05~{\rm eV})^2$.
        \label{tbl:brachingration}}
\begin{tabular}{c|c|c|c|c|c}
\hline
\hline 
&$1/R$&
\multicolumn{2}{c|}{$m_{h}=115{\rm GeV}$}&
\multicolumn{2}{c}{$m_{h}=150{\rm GeV}$}\tabularnewline
\cline{3-4} \cline{5-6} 
&(eV)&
${\rm BR}(h\rightarrow b\bar{b})$&
${\rm BR}(h\rightarrow\nu_{L}\nu_{R}^{(n)})$&
${\rm BR}(h\rightarrow b\bar{b})$&
${\rm BR}(h\rightarrow\nu_{L}\nu_{R}^{(n)})$\tabularnewline
\hline 
&$10^2$&
$2.74\times10^{-5}$&
1.0000&
$1.137\times10^{-5}$&
0.9999\tabularnewline 
$\delta=3$&$10^3$&
0.02647&
0.9642&
0.01064&
0.9360\tabularnewline
&$10^4$&
0.7196&
0.02621&
0.1639&
0.01442\tabularnewline
\hline
&
$1$&
0.59&
0.15&
0.21& 0.10\tabularnewline
$\delta=2$&
$10$&
0.74&
$2.57\times 10^{-3}$&
0.1662 &$1.06\times 10^{-3}$\tabularnewline
&
$10^2$&
0.74&
$2.58 \times 10^{-5}$&
0.1663&$1.06 \times 10^{-5}$\tabularnewline
\hline
\hline
\end{tabular}
\end{center}
\end{table}

The Higgs boson can also decay into the SM decay modes $h\rightarrow b\bar{b}$ and 
$h\rightarrow W^{*}W$. 
We compute the SM Higgs decay widths using HDECAY~\cite{Djouadi:1997yw}.
It is well known that for SM Higgs boson mass $m_h < 135$~GeV, 
the decay $h\rightarrow b\bar{b}$ dominates, whereas above 150~GeV, the dominant
decay mode is $h\rightarrow W W^{*}$. These two decay channels compete with each other 
in the intermediate mass region $135~{\rm GeV} \leq m_h \leq 150$~GeV. 
 
\begin{figure}
\begin{center}
\scalebox{0.5}{\includegraphics{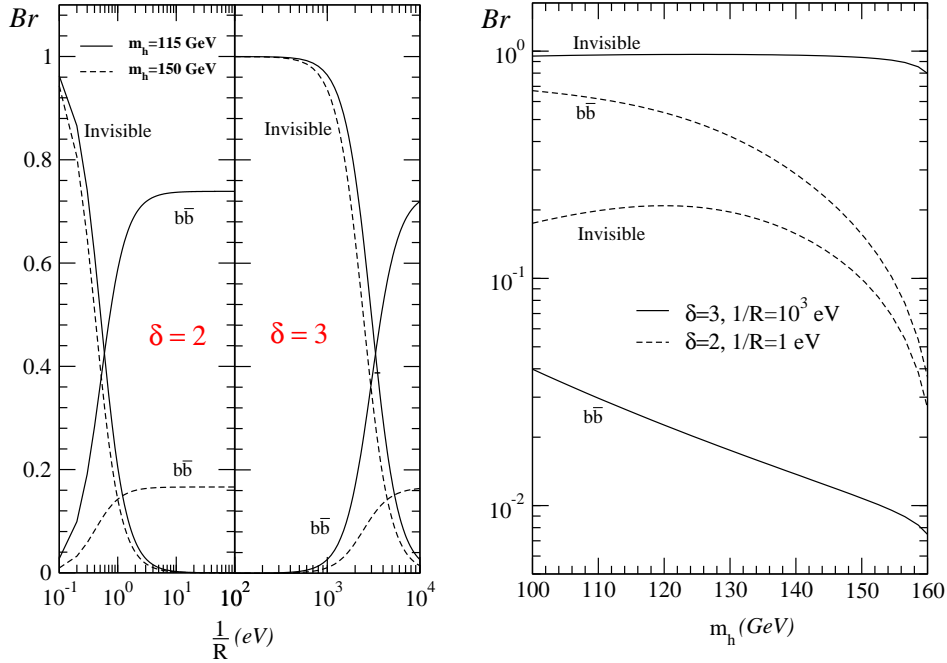}}
\caption{Decay branching ratios as a function of $1/R$ and $m_h$ for 
$\sum_{i,\ell} |\bar{m}_{\nu}^{i\ell}|^2 = (0.05~{\rm eV})^2$.
\label{fig:decay_br}}
\end{center}
\end{figure}

The decay branching ratios of ${\rm BR}(h\rightarrow b\bar{b})$ and 
${\rm BR}(h\rightarrow\nu_{L}\nu_{R}^{(n)})$ as a function of $1/R$ 
are shown in Fig.~\ref{fig:decay_br}(a) for $m_h=115$~GeV and $m_h=150$~GeV for $\delta=2,3$. 
The exact values of the branching ratios are listed in Table~\ref{tbl:brachingration}. 
Due to the enhancement from the large number of KK modes that are accessible at smaller $1/R$, 
the invisible decay mode becomes dominant as is the case, for example, for 
$\delta=3$, $1/R \lesssim 10^3$~eV. 
Fig.~\ref{fig:decay_br}(b) shows the branching ratios as a function of $1/R$ for $\delta=2,3$. 

Even though the $h\rightarrow W W^{*}$ decay mode is comparable to $h\rightarrow b\bar{b}$
for $m_h \gtrsim 150$~GeV, this mode is harder to be detected, both in the leptonic and 
hadronic decay modes of the $W W^{*}$, given the suppression in the decay branching ratio
in the former and the dominance of the background in the latter. Therefore we will not
consider the $h\rightarrow W W^{*}$ decay mode further in this study.

\subsection{Monte Carlo analysis}
In this section, we perform a Monte Carlo study to show how to detect such a signal
at the LHC, for Normal, Inverted and Degenerate mass schemes.
In order to establish the LHC range of sensitivity, we consider two options for
the integrated luminosity, $100\,{\rm fb}^{-1}$ and $1000\,{\rm fb}^{-1}$ (SLHC).
Since the signal cross section is very small for $\delta=2$ (c.f. Fig.~\ref{fig:prod_cs}), 
we will focus on $\delta=3$ in the remainder. 

As is evident from Table~\ref{tbl:brachingration}, the invisible decay mode 
$h\rightarrow\nu_{L}\nu_{R}^{(n)}$ is the dominant one for $1/R \lesssim 10^3$~eV. 
When the Higgs boson decays into the invisible mode, the produced KK state neutrino
behaves as a massive, noninteracting, stable particle and thus appears as missing energy
in the detector. Therefore, we can observe only the leptons plus $\mEt$, or
hadrons plus $\mEt$, for which there isn't enough kinematic information to reconstruct the
Higgs boson, and we can probe the signal only indirectly through the missing energy distribution.
On the contrary, in the $h\rightarrow b\bar{b}$ decay channel, we can use the $b\bar{b}$
pair to reconstruct the Higgs boson invariant mass; however, this suffers from the small decay 
branching ratio into this mode for $1/R \lesssim 10^3$~eV. In the following, 
we describe in succession the 
$h\rightarrow b\bar{b}$ and $h\rightarrow \nu_{L}\nu_{R}^{(n)}$ decay modes.

We find in agreement with Ref.~\cite{Godbole:2003it}, that for our choice of parameters, 
if only the SM Higgs production process \[ 
q\bar q\rightarrow W^* \rightarrow W(\rightarrow \ell \nu) h \]
is considered, followed by the new physics invisible 
decay of the Higgs boson $\left(h\rightarrow \nu_L \nu_R^{(n)}\right)$, 
the signal cannot be distinguished from the SM backgrounds.
However, if the new physics $\nu_R^{(n)}$-Higgs production process 
shown in Fig.~\ref{fig:signal} is also included, we will show here by performing a Monte Carlo
analysis that the signal event can be detected in certain regions of parameter space.

\subsubsection{$h\rightarrow b\bar{b}$ mode \label{sec:bb_mode}}
For the $b\bar{b}$ decay mode, we have the signal process
\bea
q\bar{q}^\prime \rightarrow W^{*}\rightarrow \ell^{+} h(\rightarrow b\bar{b}) \nu_{R}^{(n)}
\eea
for which the experimental signature is the production of $\ell^{+}b\bar{b}$
(where $\ell^+=e^+,\mu^+$), associated with a large missing transverse momentum carried
away by $\nu_{R}^{(n)}$. We do not include the $\tau$ channel due to 
the additional suppression from the branching ratio of $\tau\rightarrow \pi\nu,~\rho\nu$. 
In this study we will assume that both the $b$ and $\bar{b}$ jets in the signal event can be 
tagged with a total efficiency of $50\%$. 

\begin{figure}
\dofigs{2.5in}{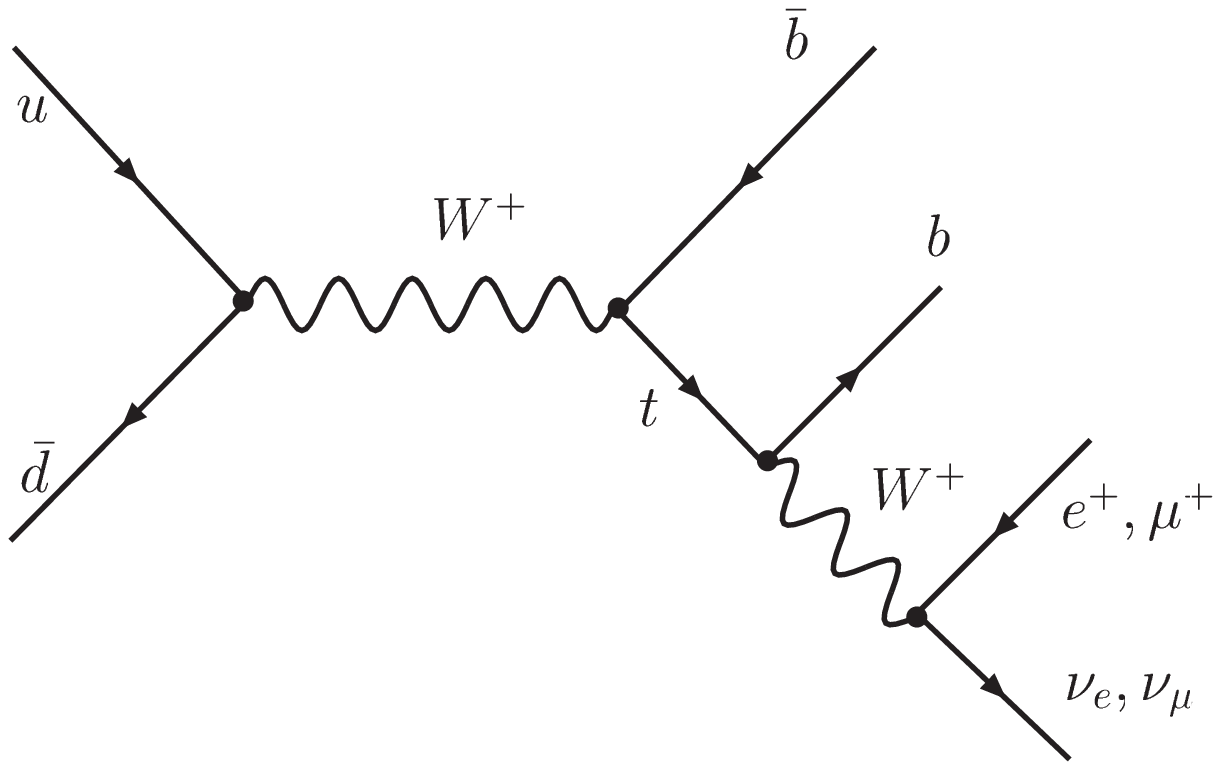}{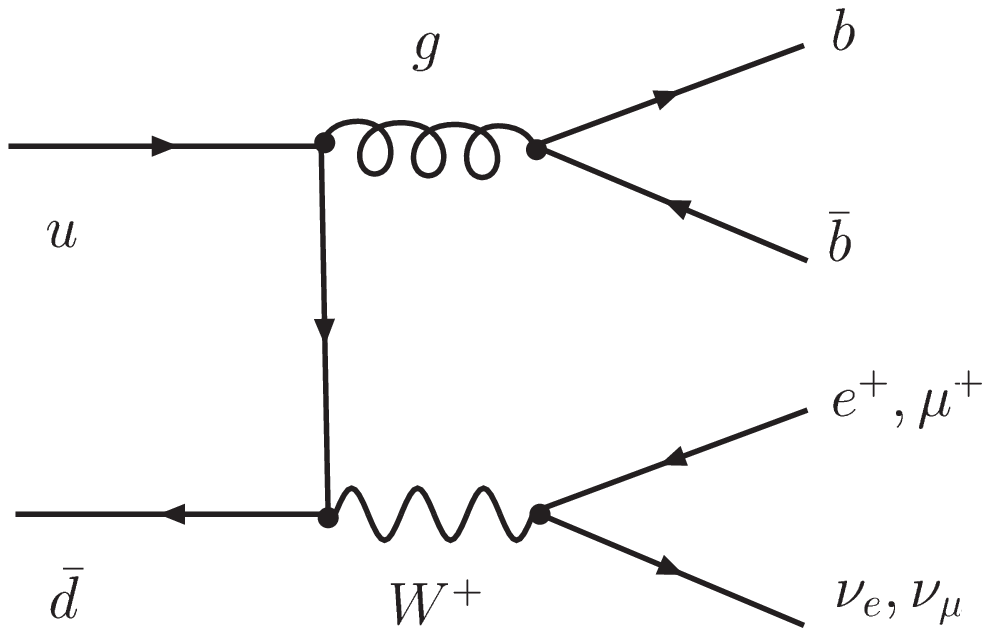}
\caption{Representative Feynman diagrams for the SM background processes \label{fig:bkgd_bb}}
\end{figure}

The intrinsic SM background processes (in addition to the $W^*\,h$ process discussed above) 
for this channel are 
\bea
t\bar{b}:\quad q\bar{q}^\prime & \rightarrow & W^{*}\rightarrow\bar{b}t(\rightarrow bW(\rightarrow \ell^{+}\nu)) \ , \\
Wb\bar{b}:\quad q\bar{q}^\prime & \rightarrow & W^{*}(\rightarrow \ell^{+}\nu)g(\rightarrow b\bar{b}),
\label{eq:bkgd}
\eea
as shown in Fig.~\ref{fig:bkgd_bb}.
In addition to the $Wb\bar{b}$ and $t\bar{b}$ processes, there are reducible SM background
processes containing misidentified or undetected particles (charged leptons or jets) which 
mimic the signal event. However, they are not expected to pose a serious problem 
after imposing kinematic cuts to veto additional jet (or lepton) activity and requiring the
event to have a large missing transverse momentum. 
We will therefore only focus on the intrinsic SM backgrounds, $Wb\bar{b}$ and $t\bar{b}$.

\begin{figure}
\includegraphics[scale=0.6]{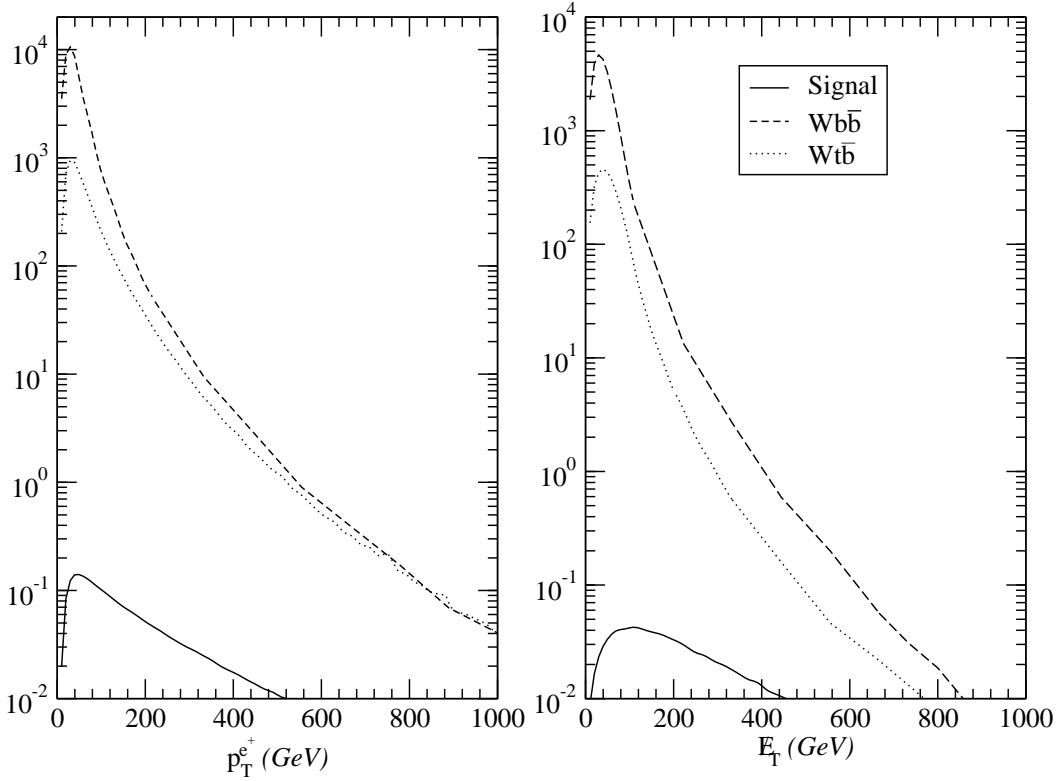}
\caption{Distribution (Number of events/GeV) of $p^{l}_{T}$, $\EmissT$, and $m^{(l\nu)}_T$
	 at LHC, in the $l^+ b \bar{b}\EmissT$ channel, for $\delta=3$,
         $m_h=115$~GeV, $1/R=500$~eV,
	 after imposing the basic cuts specified in Eq.~(\ref{eq:basic_cut}).
         \label{fig:basic_cut}}
\end{figure}

To investigate the potential of the LHC to detect such a signal, we take $1/R=500$~eV
 with $\delta=3$ and
perform a Monte Carlo simulation for $m_{h}=115~{\rm GeV}$.
To compare the relevant background event rates to the signal event
rate, we shall assume the integrated luminosity of the LHC to be $100\,{\rm fb}^{-1}$.
Here, we will assume a perfect detector that can precisely measure
the four-momenta of the final state partons. We require the separation
in $\Delta R \equiv\sqrt{(\Delta\phi)^{2}+(\Delta\eta)^{2}}$
between any two observable final state partons (not including neutrinos)
to be larger than 0.4, where $\Delta\phi$ and $\Delta\eta$ are the
separation in azimuthal angle and rapidity, respectively.
We will require the transverse momentum ($p_{T}$) and the rapidity
($\eta$) of $b$, $\bar{b}$ and $\ell^{+}$ to satisfy the following basic cuts: 
\bea
p_{T}^{q} & > & 15\, GeV,\quad\left|\eta^{q}\right|<3.0, \nonumber \\
p_{T}^{\ell} & > & 15\, GeV,\quad\left|\eta^{\ell}\right|<2.5 , \nonumber \\
\mEt      & > & 15\, GeV,\quad \Delta R  >  0.4.
\label{eq:basic_cut}
\eea
Fig.~\ref{fig:basic_cut} shows the signal and background distributions after imposing the basic cuts.
We get only 8 signal events (in all three neutrino mass schemes), and $2.5\times 10^5$ 
background events. Since this mode suffers from such a huge background, the signal events cannot be seen at the LHC.
Imposing a hard cut on the missing energy or increasing the luminosity 
(SLHC with $1000\,{\rm fb}^{-1}$), would not improve this result much.
Therefore, due to the small decay branching ratio of $h\rightarrow b \bar{b}$,
we conclude that it would be extremely difficult to use the $b\bar{b}$ mode to 
directly detect the signal at LHC or SLHC.

\subsubsection{$h\rightarrow\nu_{L}\nu_{R}^{(n)}$ (Invisible mode)\label{sec:invis_mode}}
\begin{figure}
\includegraphics[scale=0.7]{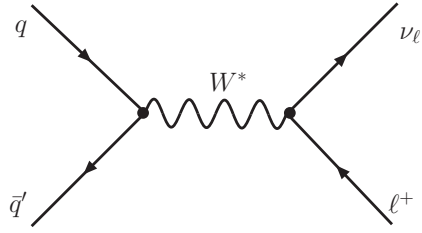}
\caption{SM Background process for invisible decay mode.\label{fig:invisible-bkgd}}
\end{figure}

For the invisible mode, we have the signal process \[
q+\bar{q}^{\prime}\rightarrow\ell^{+}+\nu_{R}+H(\rightarrow\nu\bar{\nu}_{R}^{(n)}),\]
where $\ell=e,\,\mu,\,\tau$. For the $\tau$ channel, the $\tau^+$ further decays either into
leptons or hadrons. For simplicity we only consider the $\pi^{+}\nu$ decay mode 
(with ${\rm BR}(\tau\rightarrow\pi\nu)=0.11$) of $\tau^+$ to detect the signal events. Therefore,
the collider signature for the signal process is leptons + $\mEt$ (for $\ell=e,\,\mu$), 
and $\pi^{+}+\mEt$ (for $\ell=\tau$). When $\ell=e,\mu$ we will denote this as the $\ell^{+}\mEt$
channel, but when $\ell=\tau$ we will denote this as the $\pi^{+}\mEt$ channel.
The major SM background process is the Drell-Yan charged current process \[
q\bar{q}^{\prime}\rightarrow\ell^{+}\nu \ , \]
as shown in Fig.~\ref{fig:invisible-bkgd}, where $\ell=e,\,\mu,\,\tau$.

We study next the collider signature for the normal, inverted and degenerate mass schemes. 
\paragraph{Normal Hierarchy:}
Using the Higgs interaction specified in Eqs.~(\ref{EQ4DLRM.EQ})~and~(\ref{mnlnor.EQ}), 
we perform a Monte Carlo analysis with $1/R=500$~eV and $1000$~eV, to illustrate the 
nature of a possible signal at the LHC (a $p-p$ collider with $14$~TeV center-of-mass energy). 
We consider an integrated 
luminosity of $100\,{\rm fb}^{-1}$, and impose the cuts, called Basic and Second cuts, as shown in
Table~\ref{tbl:invis-cuts}, for lepton+$\mEt$ and $\pi+\mEt$ channels.
\begin{table}
\caption{Kinematic cuts used to select events\label{tbl:invis-cuts}}
\begin{tabular}{lcccclccccl}
\hline
&
&
&
&
&
$\ell^{+}(\ell=e,\,\mu)+\mEt$&
&
&
&
&
$\pi^{+}+\mEt$\tabularnewline
\hline
Basic cuts&
&
&
&
&
$p_{T}^{\ell}>15\,{\rm GeV}$&
&
&
&
&
$p_{T}^{\pi}>15\,{\rm GeV}$\tabularnewline
&
&
&
&
&
$\mEt>15\,{\rm GeV}$&
&
&
&
&
$\mEt>15\,{\rm GeV}$\tabularnewline&
&
&
&
&
$\left|\eta^{\ell}\right|<2.5$&
&
&
&
&
$\left|\eta^{\pi}\right|<3.0$\tabularnewline
\hline
Second cuts&
&
&
&
&
$\mEt>400\,{\rm GeV}$&
&
&
&
&
$\mEt>400\,{\rm GeV}$\tabularnewline
\hline
\end{tabular}
\end{table}
\begin{table}
\caption{Number of signal ($S$) and background ($B$) events for the normal
hierarchy scheme, with $1/R=500$~eV, at the LHC with an integrated
luminosity of $100\,{\rm fb^{-1}}$. The kinematic cuts listed in
each row are applied sequentially.\label{tbl:num_normalhierachy}}
\begin{tabular}{l|c|c|c|c|c|c|c|c|c|c}
\hline 
&
\multicolumn{6}{c|}{$\ell^{+}+\mEt$}&
\multicolumn{4}{c}{$\pi^{+}+\mEt$}\tabularnewline
\hline 
&
$\ell=e$&
$\ell=\mu$&
$S$&
$B$&
$S/B$&
$S/\sqrt{B}$&
$S$&
$B$&
$S/B$&
$S/\sqrt{B}$\tabularnewline
\hline 
Basic cuts&
26&
1691&
1717&
$1.07\times10^{9}$&
$1.6\times10^{-6}$&
0.05&
191&
$0.17\times10^{9}$&
$1.12\times10^{-6}$&
0.015\tabularnewline
\hline
Second cuts&
6&
393&
399&
2432&
0.16&
8.1&
58&
27&
2.15&
11.2
\tabularnewline
\hline
\end{tabular}
\end{table}

\begin{figure}
\vspace{6mm}
\includegraphics[%
   scale=0.6]{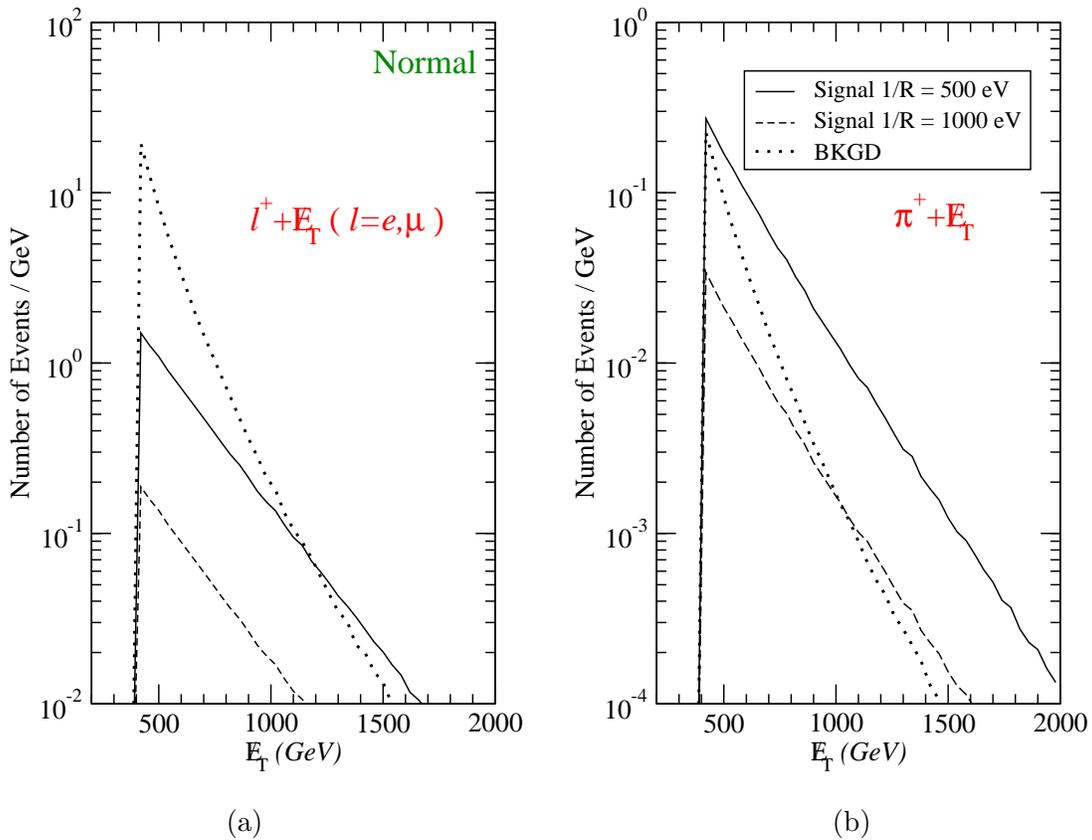}
(a)~~~~~~~~~~~~~~~~~~~~~~~~~~~~~~~~~~~~~~~~~~~~~~~~~~~~~~~~~~~(b)
\caption{Missing energy distribution (after both Basic and Second cuts) for the normal 
hierarchy scheme, with $1/R=500\,{\rm eV}$and $1000\,{\rm eV}$. 
(a) is for $\ell^{+}\mEt$ channel, and (b) for $\pi^{+}\mEt$ channel.  
\label{fig:misset_normal}}
\end{figure}

We summarize the number of the signal events ($S$) and background events ($B$), after imposing the
Second cuts, in Table~\ref{tbl:num_normalhierachy} along with the ratio of the number of signal 
to background events ($S/B$) and the statistical significance of the signal ($S/\sqrt{B}$).
In Fig.~\ref{fig:misset_normal}, we show the missing energy distribution for $1/R = 500$~eV and 
$1000$~eV, after imposing both the Basic and Second cuts, for $\ell^{+}\mEt$ in (a), 
and $\pi^{+}\mEt$ in (b). 

In the $\ell^{+}\mEt$ channel, Fig.~\ref{fig:misset_normal}(a) shows
that for $1/R=500$~eV the signal can be distinguished from background at large $\mEt$.
This is possible because the density of KK states ($d_{\hat n}$ in Eq.~(\ref{DNC.EQ})) 
increases with increasing mass of the KK state, due to which the $\mEt$ distribution tends to 
peak at higher $\mEt$ compared to the background. Even though this effect is somewhat reduced due
to the rapidly falling PDF's, the signal still tends to peak at larger $\mEt$.  

In the $\pi^{+}\mEt$ channel, as shown in Fig.~\ref{fig:misset_normal}(b), the $\tau^+$ momentum 
in the signal is harder compared to the background
since it is balanced against the heavy $\nu_R^{(n)}$-Higgs system due to which the resulting $\pi^+$ 
from the $\tau^+$ decay is also harder. Therefore after imposing the hard $\mEt$ cut (Second cuts), 
more background events will be cut away compared to the signal events leading to a significance
of $11.2\,\sigma$, as shown in Table~\ref{tbl:num_normalhierachy}. 

\begin{figure}
\vspace{6mm}
\includegraphics[%
  scale=0.6]{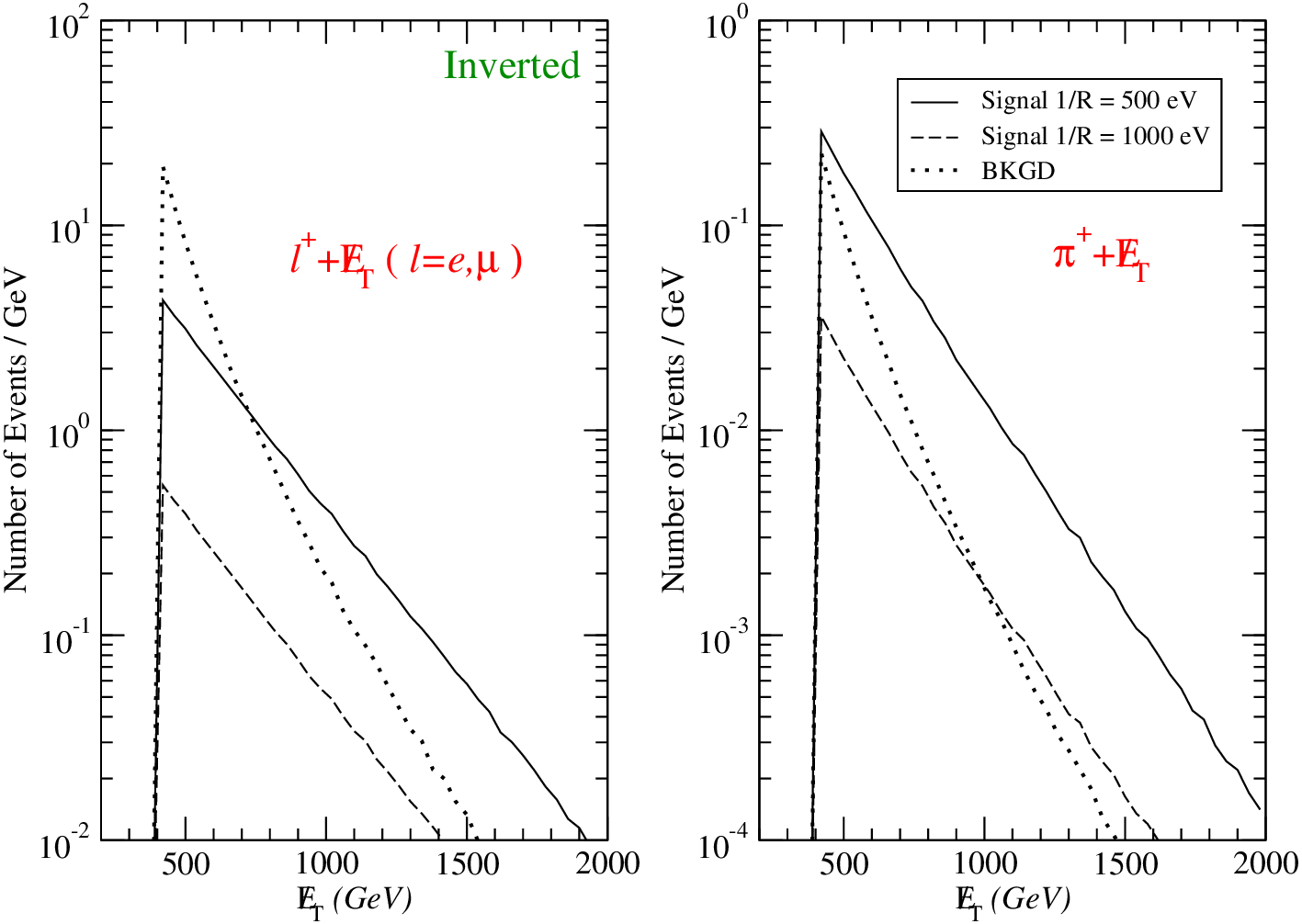}
(a)~~~~~~~~~~~~~~~~~~~~~~~~~~~~~~~~~~~~~~~~~~~~~~~~~~~~~~~~~~~(b)
\caption{Missing energy distribution (after both Basic and Second cuts) for the inverted hierarchy 
scheme, with $1/R=500\,{\rm eV}$and $1000\,{\rm eV}$. 
(a) is for $\ell^{+}\mEt$ channel, and (b) for $\pi^{+}\mEt$ channel.
\label{fig:misset_invert}}
\end{figure}

\paragraph{Inverted Hierarchy:}
We follow a similar procedure and perform a Monte Carlo using the Higgs interaction
specified in Eqs.~(\ref{EQ4DLRM.EQ})~and~(\ref{mnlinv.EQ}). We use the same kinematic cuts given
in Table~\ref{tbl:invis-cuts} for lepton+$\mEt$ and $\pi+\mEt$.
The number of signal and background events after
imposing these cuts are summarized in Table~\ref{tbl:num_inverthierachy}.
The ratio of the number of signal and background events ($S/B$), as well as the 
statistical significance of the signal ($S/\sqrt{B}$), are also shown in 
Table~\ref{tbl:num_inverthierachy}. In Fig.~\ref{fig:misset_invert} we show the missing 
energy distribution after imposing both the Basic and Second cuts, for $\ell^{+}\mEt$ in (a), 
and $\pi^{+}\mEt$ in (b).
\begin{table}
\caption{Number of signal ($S$) and background ($B$) events for the inverted
hierarchy scheme, with $1/R=500\,{\rm eV}$, at the LHC with an integrated
luminosity of $100\,{\rm fb^{-1}}$. The kinematic cuts listed in
each row are applied sequentially.\label{tbl:num_inverthierachy}}
\begin{tabular}{l|c|c|c|c|c|c|c|c|c|c}
\hline
&
\multicolumn{6}{c|}{$\ell^{+}+\mEt$}&
\multicolumn{4}{c}{$\pi^{+}+\mEt$}\tabularnewline
\hline
&
$\ell=e$&
$\ell=\mu$&
$S$&
$B$&
$S/B$&
$S/\sqrt{B}$&
$S$&
$B$&
$S/B$&
$S/\sqrt{B}$\tabularnewline
\hline
Basic cuts&
3372&
1649&
5021&
$1.07\times10^{9}$&
$4.69\times10^{-6}$&
0.15&
183&
$0.17\times10^{9}$&
$1.08\times10^{-6}$&
0.014\tabularnewline
\hline
Second cuts&
767&
375&
1142&
2432&
0.47&
23.2&
55&
27&
2.04&
10.58\tabularnewline
\hline
\end{tabular}
\end{table}
As can be seen from Table~\ref{tbl:num_inverthierachy}, for $1/R=500$~eV, there is
a significant deviation from SM background in both the $\ell^{+}\mEt$ and 
$\pi^{+}\mEt$ channels. 

We note that for both the normal and inverted mass schemes, the statistical significance 
for distinguishing the signal from background becomes marginal for $1/R \gtrsim 1000$~eV, 
due to an overall suppression of the signal rate in the entire $\mEt$ range, as
suggested by Figs.~(\ref{fig:misset_normal})~and~(\ref{fig:misset_invert}).
 
\begin{figure}
\vspace{6mm}
\includegraphics[scale=0.6]{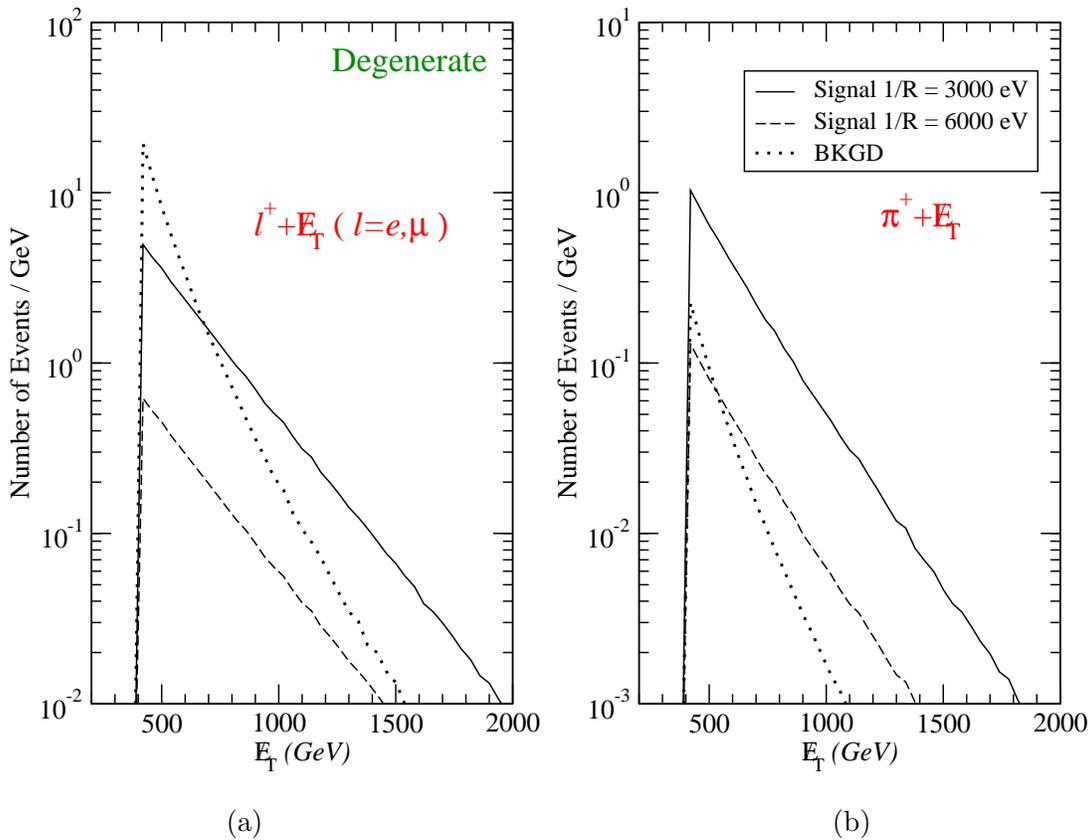}
(a)~~~~~~~~~~~~~~~~~~~~~~~~~~~~~~~~~~~~~~~~~~~~~~~~~~~~~~~~~~~(b)
\caption{Missing energy distribution (after both Basic and Second cuts) for the degenerate 
mass scheme, with $1/R=3000\,{\rm eV}$ and $6000\,{\rm eV}$. 
(a) is for $\ell^{+}\mEt$ channel, and (b) for $\pi^{+}\mEt$ channel.
\label{fig:misset_degen}}
\end{figure}

\paragraph{Degenerate:}
As already pointed out, we expect the constraints on $1/R$ from neutrino oscillation and
unitarity to be stronger in the degenerate mass scheme, and we therefore choose to present 
results for $1/R=3000$~eV, as this yields the same
order of magnitude total cross section as in the normal and inverted mass schemes.
We perform a Monte Carlo study with the Higgs interaction given by
Eqs.~(\ref{EQ4DLRM.EQ})~and~(\ref{mnldeg.EQ}).
The numerical results are summarized in Table.~\ref{tbl:num_degeneratehierachy} and 
Fig.~\ref{fig:misset_degen}, and in the latter we also show the distributions for $1/R=6000$~eV. 
\begin{table}
\caption{Number of signal ($S$) and background ($B$) events for the degenerate
mass scheme, with $1/R=3000\,{\rm eV}$, at the LHC with an integrated
luminosity of $100\,{\rm fb^{-1}}$. The kinematic cuts listed in
each row are applied sequentially.\label{tbl:num_degeneratehierachy}}
\begin{tabular}{l|c|c|c|c|c|c|c|c|c|c}
\hline
&
\multicolumn{6}{c|}{$\ell^{+}+\mEt$}&
\multicolumn{4}{c}{$\pi^{+}+\mEt$}\tabularnewline
\hline
&
$\ell=e$&
$\ell=\mu$&
$S$&
$B$&
$S/B$&
$S/\sqrt{B}$&
$S$&
$B$&
$S/B$&
$S/\sqrt{B}$\tabularnewline
\hline
Basic cuts&
6049&
6065&
12114&
$1.07\times10^{9}$&
$1.1\times10^{-5}$&
0.37&
672&
$0.17\times10^{9}$&
$4.0\times10^{-6}$&
0.05\tabularnewline
\hline
Second cuts&
1376&
1380&
2756&
2432&
1.13&
55.9&
202&
27&
7.48&
38.87\tabularnewline
\hline
\end{tabular}
\end{table}
In the degenerate mass scheme, there are about the same number of signal events in the 
$e^+$, $\mu^+$ and $\tau^+$ channels. 

\subsection{Discovery potential}
\begin{figure}
\includegraphics[scale=0.6]{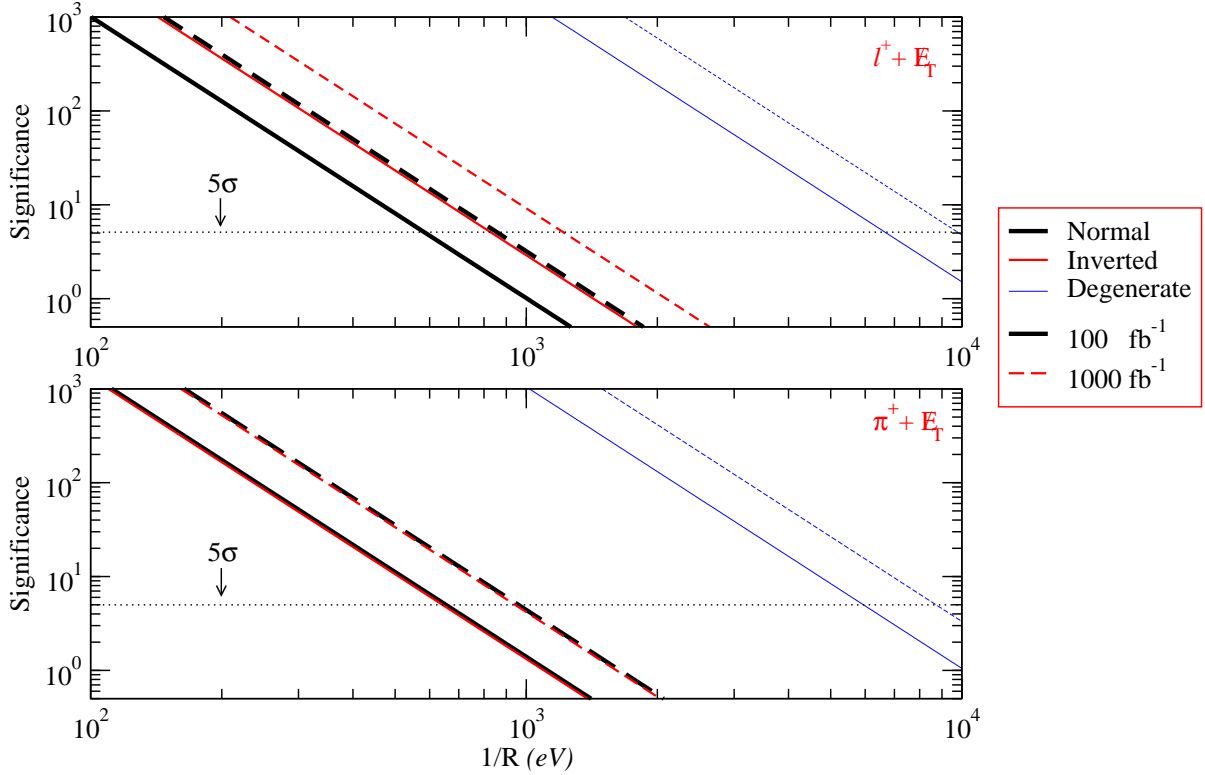}
\caption{Significance (after Basic and Second cuts) as a function of $1/R$ ($\delta=3$) 
for the normal, inverted and degenerate 
mass schemes at LHC with an integrated luminosity ${\mathcal L}=100$ and $1000~ {\rm  fb}^{-1}$.
\label{fig:significance}}
\end{figure}

In order to study the potential of the LHC to distinguish the signal from background events, 
we compute 
the Significance
\beq
S_B = \frac{S}{\sqrt{B}} = \frac{\sigma_{signal}{\mathcal L}}{\sqrt{\sigma_{bkgd}{\mathcal L}}},
\eeq
where ${\mathcal L}$ is the integrated luminosity. To unambiguously establish the new physics
scenario we are considering, it is desirable to have $S_B>5$. 
For each mass scheme, we plot in Fig.~\ref{fig:significance}, $S_B$ as a function of $1/R$, 
at the LHC for ${\mathcal L}=100$ and $1000~{\rm fb^{-1}}$. In this, we have imposed both Basic and 
Second cuts listed in Table.~\ref{tbl:invis-cuts} for the $\ell^+\mEt$ and $\pi^+\mEt$ channels.

We infer from Fig.~\ref{fig:significance} that the $\ell^+\mEt$ and $\pi^+\mEt$ channels
lead to a similar significance, with $\ell^+\mEt$ being somewhat better in the
case of the inverted mass scheme.  
If $1/R$ is about 900~eV (for normal or inverted) or about 8~KeV (for degenerate) we may 
expect $S_B$ in the $2-5\,\sigma$ range at the LHC. Here, it is worth pointing out that if we 
strictly impose the unitarity and oscillation constraints on $1/R$ that we derived in 
Ref.~\cite{Cao:2003yx}, we would expect a poor $S_B$ at the LHC with ${\mathcal L}=100~{fb}^{-1}$. 
However, we had pointed out that owing to the dependence 
on the cutoff for $\delta=3$, those constraints are uncertain to some extent. Therefore, 
in the event that the constraint on $1/R$ is relaxed somewhat, we may expect to see a 
signal at the LHC. 

The signal and background event rate estimates discussed thus far are
subject to PDF uncertainties~\cite{Martin:1999ww,Stump:2003yu}.
Since both our signal and background processes are mediated through a virtual
$W^*$ production, we can estimate this uncertainty
by scanning over the 41 sets of CTEQ6.1 parton distribution functions~\cite{Stump:2003yu}
to calculate the next-to-leading order $W^*$ production cross section.
We find that the uncertainty in $W^*$ (Drell-Yan) production, over the $Q^2$ region of interest, 
is about $2\,\%$ at the LHC energy.
Since both our signal and background processes go through a $W^*$,
this uncertainty largely cancels and yields about the same $S/B$ value.
Furthermore, when data becomes available at the LHC, the PDF uncertainties will be further 
reduced from global analysis.
Another way is to use the side-band method in order to estimate the SM
background rate in the signal region, enabling us to minimize
the PDF uncertainties.
For example, in our case we could use the background dominated $\mEt < 400$~GeV region 
as the side-band to estimate the SM background rate.

\subsection{Distinguishing mass schemes}
We pointed out in Sec.~\ref{BNLED_HINT.SEC} that the neutrino oscillation data presently leave
undetermined the sign of $\Delta m^2_{\rm atm}$, in addition to the absolute scale of neutrino
masses. Measuring the sign of $\Delta m^2_{\rm atm}$ determines directly 
whether the normal or the inverted hierarchy is realized in nature. While presently no experiment
is capable of achieving this, the proposed long baseline neutrino experiments could potentially
 determine this sign by making use of matter effects in the earth~\cite{Barger:1999jj}. 
Here, it is interesting to ask if we can 
instead use the collider observables that we have been considering in order to 
distinguish between the normal, inverted and the degenerate mass schemes, 
given an excess above SM background at the LHC that is consistent with
bulk right handed neutrinos as is being discussed here. 
We define ${\mathcal N}(\mu + \mEt)$ and ${\mathcal N}(e + \mEt)$ to be the number of $(\mu + \mEt)$ 
and $(e + \mEt)$ signal events, respectively, after the Second cut.
As a first step, we assume that  ${\mathcal N}(\mu + \mEt)$ and ${\mathcal N}(e + \mEt)$ can be determined 
by subtracting out the estimated number of background $(\mu + \mEt)$ and $(e + \mEt)$ events, from the actual 
number of events seen in the collider.
We find a suitable discriminant that could potentially 
distinguish between the neutrino mass schemes to be
\beq
\mathcal{A}_{\mu e} \equiv \frac{{\mathcal N}(\mu+\mEt) - {\mathcal N}(e+\mEt)}{{\mathcal N}(\mu+\mEt) + {\mathcal N}(e+\mEt)} \ .
\eeq
It can be shown from Eqs.~(\ref{DMSQ.EQ}),~(\ref{EQ4DLRM.EQ})~and~(\ref{eq:hadron_cs}) that
\bea
{\mathcal N}(e+\mEt) &\propto& \frac{\mathcal L}{(1/R)^\delta} \left[ m_1^2 + (0.531)^2 \Delta m^2_{solar} \right] \ , \nonumber \\
{\mathcal N}(\mu+\mEt) &\propto& \frac{\mathcal L}{(1/R)^\delta} \left[ m_1^2 + (0.599)^2 \Delta m^2_{solar} \pm (0.707)^2 \Delta m^2_{atm} \right] \ ,
\label{NUMEMU.EQ}
\eea
where the upper (lower) sign is for the normal (inverted) hierarchy.\footnote{
For the Second cuts with $\delta=3$, we find, say from Table~\ref{tbl:num_degeneratehierachy}, 
the constant of proportionality in Eq.~(\ref{NUMEMU.EQ}) to be $3.726\times 10^{11}$~eV~fb.}
Therefore,
\beq
\mathcal{A}_{\mu e} \approx \frac{\pm 0.5 \, \Delta m^2_{atm}}{2 m_1^2 \pm 0.5 \, \Delta m^2_{atm}} \ ,
\label{Amue.EQ}
\eeq
and for normal (inverted) hierarchy $\mathcal{A}_{\mu e} > 0$ ($\mathcal{A}_{\mu e} < 0$). 
\begin{figure}
\dofig{5in}{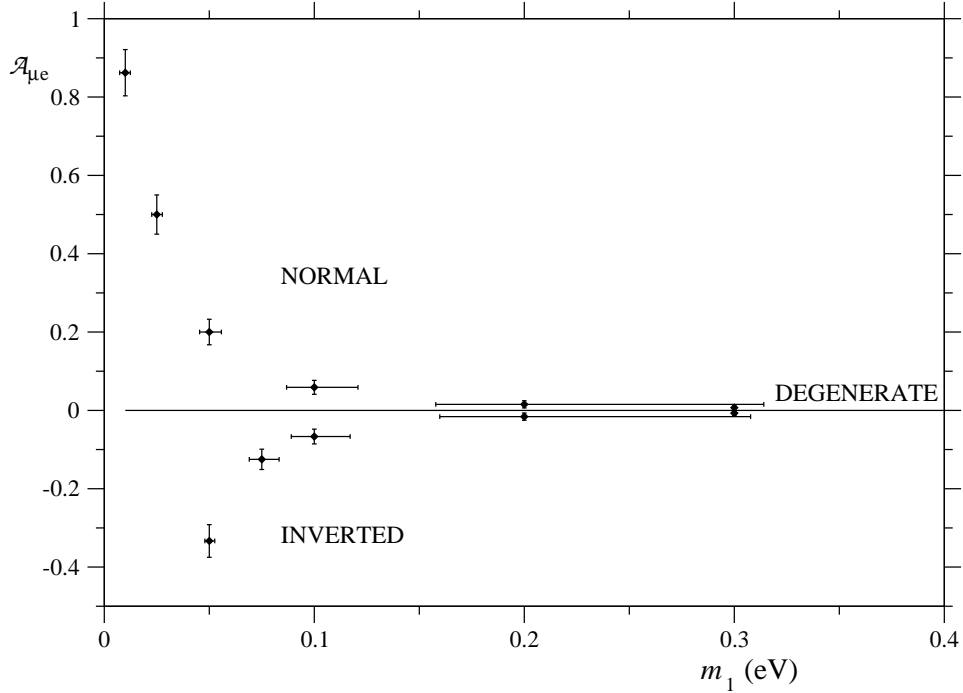}
\caption{$\mathcal{A}_{\mu e}$ as a function of $m_1$. $\mathcal{A}_{\mu e} > 0$ ($\mathcal{A}_{\mu e} < 0 $) implies normal (inverted) hierarchy and $\mathcal{A}_{\mu e} \approx 0$ implies degenerate mass scheme. The error-bars are for the LHC with $\mathcal{L}=100~{\rm fb}^{-1}$ and $1/R=500$~eV.
\label{Amuem1.FIG}}
\end{figure}

In Fig.~\ref{Amuem1.FIG} we plot $\mathcal{A}_{\mu e}$ as a function of $m_1$.
We note here that for the inverted mass scheme, the smallest value that $m_1$ can take is $0.05$~eV, 
as can be seen from Fig.~\ref{MNUSCH.EQ}. 
We see from Fig.~\ref{Amuem1.FIG} 
that it is indeed possible to distinguish between the normal and inverted hierarchies. 
The vertical error-bar shows $\delta \mathcal{A}_{\mu e}$, the statistical accuracy with which
$\mathcal{A}_{\mu e}$ can be measured at the LHC, for a luminosity $\mathcal{L}=100~{\rm fb}^{-1}$. 
We estimate $\delta \mathcal{A}_{\mu e}$ by calculating the number of $e$ and $\mu$ events 
using Eq.~(\ref{NUMEMU.EQ}). For $m_1$ small, we see that there is excellent discriminating power of 
many standard deviations to determine whether the normal or the inverted mass hierarchy is realized. 
However, as $m_1$ increases the number of $e,~\mu~{\rm and}~\tau$ events become approximately equal
and the discriminating power diminishes. 

The absolute scale of neutrino masses (say $m_1$) can also be determined by measuring 
$\mathcal{A}_{\mu e}$, but only if $m_1^2$ is not too large compared to $\Delta m^2$. 
This is because, a given (statistical) uncertainty 
$\delta \mathcal{A}_{\mu e}$ corresponds to an uncertainty on the mass given by 
$\delta m_1^2 = (0.25 \Delta m^2_{atm}) (\delta \mathcal{A}_{\mu e}/\mathcal{A}_{\mu e}^2)$. 
Thus, for $m_1^2 \gg \Delta m^2$ the masses become approximately 
degenerate, ${\mathcal N}(e+\mEt)\approx {\mathcal N}(\mu+\mEt) \approx {\mathcal N}(\tau+\mEt)$, due to which 
$\mathcal{A}_{\mu e} \approx 0$, and the uncertainty $\delta m_1^2$ becomes worse. 
From Fig.~\ref{Amuem1.FIG} we see that we can determine $m_1$ by measuring $\mathcal{A}_{\mu e}$, 
and the horizontal error-bars shows the uncertainty in the inferred value of $m_1$ due to the 
(statistical) uncertainty $\delta \mathcal{A}_{\mu e}$ on the measured $\mathcal{A}_{\mu e}$. 
This might provide a means to measure the absolute scale of neutrino mass ($m_1$) 
which cannot be done through neutrino oscillation experiments.  
For example, in the normal mass scheme if $m_1=0.05$~eV, at the LHC we would measure an asymmetry
$\mathcal{A}_{\mu e} = 0.2 \pm 0.03$, where the error is the statistical uncertainty
for $\mathcal{L}=100~{\rm fb}^{-1}$. From this measurement, we infer from 
Eq.~(\ref{Amue.EQ}) that $m_1 = 0.05^{+0.006}_{-0.005}$~eV, and this error is shown in Fig.~\ref{Amuem1.FIG}
as the horizontal error-bar.   
For $m_1 > 0.2$~eV the horizontal error-bars get progressively larger and we do not show them 
in the figure to reduce clutter. 
If $m_1$ is indeed large in nature, it follows from Eq.~(\ref{SIGSCA.EQ}) that from the 
measured bulk neutrino production cross section, we can at best only determine the ratio 
$m^2/(1/R)^\delta$ using Eq.~(\ref{eq:hadron_cs}). 

It is interesting to note that the neutrino
oscillation probability from an active species to the (sterile) heavier KK modes is also 
proportional to $m^2/(1/R)^\delta$~\cite{Cao:2003yx}. Therefore, it appears that in 
the degenerate mass case, it would not be possible to disentangle $m$ and $1/R$ even if we 
observed the collider signature that we have been considering; this would still be the 
case if we also observed a finite oscillation probability into the heavier KK states.

In Eq.~(\ref{lBFIT.EQ}), for simplicity, we assumed $l^{e3}$ to be zero. If we had not done so, 
one might wonder if $\mathcal{A}_{\mu e}$ has any sensitivity to the small angle associated 
with $l^{e3}$. We do not expect this to be measurable since the change in the numbers of $e$ and
$\mu$ events would be much smaller than the statistical uncertainties involved. 
It is also worth mentioning that the $\tau$-$e$ (or $\tau$-$\mu$) asymmetry does not add any 
information in probing the neutrino masses. 

\section{Conclusions}
\label{CONCL.SEC}
We consider a theory with right-handed neutrinos propagating in $\delta$ large
extra dimensions, with radius $R$, and coupled to the left-handed neutrino by the Higgs boson 
Yukawa coupling. Such a theory naturally explains the smallness of the neutrino mass scale,
in addition to addressing the gauge hierarchy problem in the standard model (SM).
Presently the combination of neutrino oscillation data leave undetermined the absolute 
neutrino mass scale $m_\nu$ which lead us to present our results for the normal, inverted 
and degenerate mass schemes. In addition, the observables that we analyze depend on
$\delta$ and $1/R$ of the extra-dimensional theory. We take the fundamental
scale of gravity $M_*$ to be the electroweak scale $M_{EW}\sim 10^3~{\rm GeV}$. 

We showed in our previous work~\cite{Cao:2003yx} that neutrino oscillation and unitarity
in Higgs-Higgs scattering places a lower bound on $1/R$, particularly strong for $\delta > 3$.
We thus restricted our consideration to theories with $\delta \leq 3$,  
and we pointed out that the bounds we derived were somewhat uncertain for $\delta = 2~{\rm and}~3$, 
since the observables depended on the cutoff ($M_*$) of the theory. 
Here, we present our results choosing $1/R$ to be around this lower bound, although not 
imposing the bound strictly due to the uncertainty just mentioned. 

We show that the Higgs boson production at a collider and its decay can be 
enhanced significantly owing to the large number of Kaluza-Klein (KK) neutrino states that can 
be produced in the final state. To probe this theory we consider the signal process \[
q\bar{q}'\rightarrow W^{*}\rightarrow \ell^{+}\, h\, \nu_{R}^{\prime i (n)} \]
at hadron colliders, such as the Tevatron and the LHC. The production cross section
is shown in Fig.~(\ref{fig:prod_cs}) for $\delta=2,3$. We find for instance, for $\delta=3$, 
$1/R \sim 1$~KeV, we could detect a signal at the LHC with a luminosity ${\mathcal L}=100~{fb}^{-1}$.  

In addition to the SM decay mode of the Higgs boson $h\rightarrow b\bar{b}$, we also consider the 
new invisible decay mode $(h\rightarrow \nu_{L}\nu_{R}^{(n)})$. 
As shown in Fig.~(\ref{fig:decay_br}), the invisible decay mode dominates for smaller $1/R$. 

We perform a Monte Carlo analysis of the signal process along with the SM backgrounds,
and find that the Higgs boson decaying invisibly, leading to a
$\ell^+~\mEt~(\ell=e,\mu,\tau)$ signature, is more promising compared to the $h\rightarrow b\bar{b}$ 
decay mode. We apply the cuts shown in Table~\ref{tbl:invis-cuts} to enhance the signal relative to the 
background, and show the resulting significance in Fig.~(\ref{fig:significance}). We find a
significance in the $2-5~\sigma$ range for $\delta=3$ and $1/R \sim 900$~eV for normal and inverted 
mass hierarchies, and $1/R \sim 8000$~eV for the degenerate mass scheme.
We only consider the positive charged lepton and the $\pi^+\,\nu$\ decay mode of the $\tau^+$.
Including the negative charged lepton and the $\rho\,\nu$\ decay mode of the $\tau$ will improve
the situation somewhat.

Finally, we point out that if a positive signal is found that is compatible with the extra-dimensional
hypothesis considered here, the asymmetry in the number of $\mu$ versus $e$ events can be used to 
distinguish between the neutrino mass schemes and to determine the absolute neutrino mass scale, and
the accuracies with which these can be determined are shown in Fig.~(\ref{Amuem1.FIG}). 
The collider might be a unique place to determine this information as it is not available 
from neutrino oscillation experiments. 

\vspace*{2mm}
\noindent
{\bf Acknowledgments:}\\
We thank J.~Huston, J.~Linnemann, K.~Tobe and J.~Wells for useful discussions. 
This work was supported in part by the NSF grant PHY-0244919.


\end{document}